\documentclass{emulateapj}
\usepackage{color}
\usepackage{rotating}
\usepackage{gensymb}

 \newcommand{\vecv}{\mbox{\boldmath $v$} {}}
\newcommand{\vecV}{\mbox{\boldmath $V$} {}}
\newcommand{\vecr}{\mbox{\boldmath $r$} {}}

\newcommand{\vece}{\mbox{\boldmath $e$} {}}
\newcommand{\vecF}{\mbox{\boldmath $F$} {}}

\newcommand{\vecnabla}{\mbox{\boldmath $\nabla$} {}}
\newcommand\Rtilde{\stackrel{\sim}{\smash{R}\rule{0pt}{1.1ex}}}
\usepackage{epsfig}
\usepackage{amsmath}
\newif\ifAMStwofonts
\shorttitle{Eccentric companions: Testing the local approximation}
\shortauthors{S\'anchez-Salcedo}
\begin{document}

\title{Orbital evolution of eccentric low-mass companions embedded in gaseous disks: testing the local
approximation}
\author{
  F. J. S\'anchez-Salcedo\altaffilmark{1} 
 }

\altaffiltext{1}{Instituto de Astronom\'{\i}a, Universidad Nacional Aut\'onoma de M\'exico, A. P. 70-264,
Mexico City 04510, Mexico \email{(jsanchez@astro.unam.mx)}}

\begin{abstract}

We study the tidal interaction between a low-mass companion (e.g., a protoplanet or a
black hole) in
orbit about a central mass, and the accretion disk within which it is submerged.
We present results for a companion on a coplanar orbit with eccentricity $e$
between $0.1$ and $0.6$. For these eccentricities, dynamical friction arguments
in its local approximation, that is, ignoring differential rotation and the curvature of the orbit, 
provide simple analytical expressions
for the rates of energy and angular momentum exchange between the disk and the companion. 
We examine the range of validity of the dynamical friction approach by conducting
a series of hydrodynamical simulations of a perturber with softening radius $R_{\rm soft}$
embedded in a two-dimensional disk. We find close agreement between predictions and the
values in simulations provided that $R_{\rm soft}$ is chosen sufficiently small, below
a threshold value $\Rtilde_{\rm soft}$, which depends on the disk parameters and on $e$.
We give $\Rtilde_{\rm soft}$ for both razor-thin disks and disks with
a finite scaleheight. For point-like perturbers, the local
approximation is valid if the accretion radius is smaller than $\Rtilde_{\rm soft}$.
This condition imposes an upper value on the mass of the perturber.

\end{abstract}


\section{Introduction}
\label{sec:intro}

There are numerous studies about the tidal interaction between a disk in Keplerian rotation about a 
central mass and a low-mass companion. Determining the orbital evolution of the companion is 
crucial to understand a range of astrophysical scenarios. Embryos, protoplanetary
cores and planets change its orbital parameters (semimajor axis $a$, eccentricity $e$
and inclination $i$) due to the mutual gravitational scatterings and due to the
exchange of angular momentum and energy with the protoplanetary
disk \citep[e.g.,][]{bar14}. Likewise, stars, stellar black holes and other compact objects 
experience orbital evolution within accretion disks around supermassive black holes in active
galactic nuclei \citep[e.g.,][]{koc11}.

In this paper we are interested in the interaction between the disk and a companion in
an eccentric and coplanar orbit with $e>0.1$. A substantial body of research has been directed to quantify
the orbital evolution of eccentric perturbers through semianalytical models 
\citep[e.g.,][]{gol80,art94,pap00,gol03,tan04,mut11}.
or using numerical simulations \citep[e.g.,][]{cre06,cre07,mar09,bit10,bit11,bit13,fen14,duf15,rag18}.
For perturbers with such a small mass that they have a weak impact on the disk, 
these studies show that the response of the disk depends 
on the parameter $X\equiv e/h$, where $h$ is the aspect ratio of the disk (typically $h=0.04$).
For small $X$, the perturber 
describes epicyclic motions of small amplitude, and it excites a trailing and a leading spiral
wave, because of the Keplerian shear of the flow in its vicinity \citep[e.g.,][]{tan04}.
As $X$ is raised, the mean velocity of the perturber relative to the local gas increases, and 
therefore the shear becomes less important. 
For instance, in the simulations of \citet{cre07} with $X=6$, a significant density enhancement 
appears in front of the perturber when it is at apocenter, whereas 
the enhancement lags behind it at pericenter.

The eccentricity distribution of exoplanets is broad, with a median value around
$0.3$ \citep[e.g.,][]{mar05,udr07,xie16,mil19}.
Some extrasolar planets have eccentricities larger than $0.6$ \citep{wit07,tam08}.
Motivated by these findings, we are concerned with the 
orbital evolution of a perturber having $X\gtrsim 3$. Theoretical predictions in this regime are scarce.
\citet{pap00} evaluate the torque acting on a low-mass perturber with
$X\lesssim 5$, by including all Lindblad resonances required for convergence. 
The perturber was modeled using a softening radius between $0.4H$ and $H$, where $H$ is
the scaleheight of the disk.
For a disk with an initial surface density $\propto R^{-3/2}$ and a perturber with $X>1.1$, 
they find that the torque on the perturber is positive,  and the eccentricity is 
damped in a timescale $\propto e^{3}$.
They also note that the torque is rather sensitive to the 
softening radius. Consequently, the
ambiguity in the definition of the softening radius to be used in real three-dimensional
(3D) disks for eccentric orbits
leads to an uncertainty in the magnitude of the torque. A 3D treatment of the wake excited
within $H$ from the perturber is desirable because the wake at distances within $H$ from
the perturber contributes to the torque.
Another limitation of the resonance method is that it becomes impractical
for arbitrary large $X$, say $X>5$, because the convergence is very slow.

For $X\gtrsim 3$, \citet{mut11} suggest that a dynamical friction approach may be a good
approximation to estimate the migration timescale $\tau_{a}\equiv a/\dot{a}$ and 
the eccentricity damping timescale $\tau_{e}\equiv e/\dot{e}$ as the perturber moves 
supersonically relative to the
local gas and the Keplerian shear is unimportant \citep[see also][]{pap02,rei12}.
More specifically,
they compute the force that the disk exerts on the perturber in the local approximation, that is, 
taking the local values of the disk surface density and sound speed at each point of the orbit,
and evaluating the drag force as if the
disk were homogeneous and the orbit rectilinear\footnote{With our definition of local approximation,
a dynamical friction approach is not necessarily a local approximation. We can study the dynamical
friction force incorporating curvature terms, i.e. nonlocal effects \citep[e.g.,][]{san01,kim07},
or density gradients \citep{jus05}.}. Using this approximation, 
\citet{mut11} were able to predict $\tau_{a}$ and $\tau_{e}$ for a variety of disk models in a 
rather straightforward way. 

Another virtue of the local approximation is that the formalism can be extended to include the
vertical extent of the disk.
In fact, \citet{can13} derive the drag force exerted on a perturber moving in rectilinear orbit in
the midplane of a vertically-stratified slab. Thus, for those model parameters for which the local approximation
is confirmed to be satisfactory in 2D models, we can apply the analytical expressions
in \citet{can13}
to evaluate $\tau_{a}$ and $\tau_{e}$, following the same approach as \citet{mut11}, but
now including properly the 3D structure of the wake, which is important for supersonic
perturbers. Therefore it is essential to determine under 
which conditions the local approximation provides accurate results. To do so, we have carried out a
set of 2D numerical simulations and performed a detailed comparison between numerical results
and analytical predictions. 

The paper is organized as follows. In Section \ref{sec:model_description}, we describe the model 
and provide some relevant system quantities that
characterize the tidal interaction between the disk and the satellite.
Section \ref{sec:local_approximation} gives an overview of the dynamical friction approach in 
its local approximation. In Section \ref{sec:experiments}, we compare the results of direct numerical 
simulations with the theoretical estimates based on the local approximation. Extensions to a 3D disk and to 
point-like perturbers are discussed in Section \ref{sec:3Dapprox}. Finally, our 
findings ae summarized in Section \ref{sec:conclusions}.

\section{Model description}
\label{sec:model_description}
We consider a perturber (the companion) of mass $M_{p}$ in orbit around a central mass $M_{c}$
in the midplane of the accretion disk (i.e.~coplanar orbit).
We will asume that the ratio between masses, $q\equiv M_{p}/M_{c}$, 
is low enough that it cannot open a gap in the disk (i.e. type I migration in the 
terminology of planetary migration) and the perturbation induced in the disk is weak. 
The mass threshold $q_{\rm crit}$ to open a gap 
depends on the eccentricity; it increases as eccentricity increases \citep{hos07}.
As a guide number,  $q_{\rm crit}\simeq 10^{-3}$ for a perturber with $e=0.15$ 
embedded in a disk with a viscosity typical for protoplanetary disks.
In this paper we will consider only $q<q_{\rm crit}$. 
In the limit of low mass, the timescales of migration and eccentricity damping will be much
longer than the orbital period and, thus, we may use the osculating elements to
describe the orbital evolution of the perturber.

The total force on the perturber is $\vecF_{t}=\vecF_{0}+\vecF_{1}$, where $\vecF_{0}$ is the
gravitational force created by the central mass plus the unperturbed disk, and $\vecF_{1}$ is
the backreaction force due to the density perturbations induced in the disk. In order to calculate
the change rates of $a$ and $e$, we need the two components of $\vecF_{1}$ or, equivalently,
the power ${\mathcal{P}}_{1}$ and the torque $T_{1}$ exerted on the perturber by the
density wake excited in the disk. 
Along the paper, we will use the convention that $T_{1}$ is negative when the perturber 
loses angular momentum, and it is positive otherwise.
The time derivatives of $a$ and $e$ can be computed using the Gauss equations as 
\begin{equation}
\frac{da}{dt}=\frac{2{\mathcal{P}}_{1}}{a\omega^{2}M_{p}},
\label{eq:gauss1}
\end{equation}
and
\begin{equation}
\frac{de}{dt}= \frac{\eta^{2}}{e a^{2} \omega^{2}M_{p}} \left(\mathcal{P}_{1}-\frac{\omega T_{1}}{\eta}\right).
\label{eq:gauss2}
\end{equation}
where $\omega=\sqrt{GM_{c}/a^{3}}$ and $\eta=\sqrt{1-e^{2}}$.
Note that the force component $\vecF_{0}$ cannot lead to a 
net radial migration or eccentricity damping.

As it will become clear later, it is useful to compute the velocity of the 
gas relative to the perturber.
More specifically, we define the relative velocity as $\vecV_{\rm rel}=\vecv_{g}-\vecv_{p}$,
where $\vecv_{p}$ is the perturber's velocity, and $\vecv_{g}$ is the unperturbed velocity of the gas 
evaluated at the location of the perturber.
Without loss of generality, we adopt a system of reference where the perturber 
has its pericenter at $x=(1-e)a$, $y=0$ and $z=0$. 
Using a polar coordinate system $(R,\theta)$ centered on the central mass,
the velocity of the perturber is
\begin{equation}
\vecv_{p}=\frac{a\omega}{\sqrt{1-e^{2}}}\left(e\sin\theta \hat{\vece}_{r} +
[1+e\cos\theta]\hat{\vece}_{\theta}\right).
\label{eq:vel_perturber}
\end{equation}
In our system of reference, $\theta$ corresponds to the true anomaly $f$.

On the other hand, the unperturbed velocity of the gas is
\begin{equation}
\vecv_{g}(R)=R\Omega \sqrt{1+\frac{1}{\Sigma R \Omega^{2}} \frac{dP}{dR}} \hat{\vece}_{\theta},
\label{eq:vel_kepler}
\end{equation}
where $\Omega$ is the Keplerian angular velocity
$\Omega (R)=\sqrt{GM_{c}/R^{3}}$, and $P$ the unperturbed gas pressure.
From Eqs. (\ref{eq:vel_perturber}) and (\ref{eq:vel_kepler}), we can obtain $\vecV_{\rm rel}$.
Note that both the disk and the secondary rotate in the counterclockwise direction.

\begin{figure}
\includegraphics[width=92mm,height=80mm]{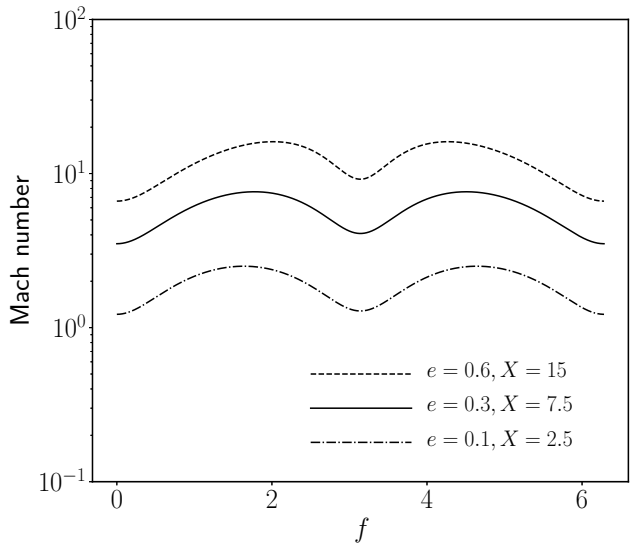}
  \caption{Local Mach number versus true anomaly for a perturber
in a Keplerian orbit with eccentricity $e$ embedded in a disk with constant aspect ratio $h=0.04$.
 }
\vskip 0.75cm
\label{fig:Mach_f}
\end{figure}

We define the local Mach number  ${\mathcal{M}}$ as $V_{\rm rel}/c_{s,p}$, where $c_{s,p}$ is the
disk sound speed at the position of the perturber.
Figure \ref{fig:Mach_f} shows ${\mathcal{M}}$ for a disk
with constant aspect ratio ($h=c_{s}/[\Omega R]=0.04$), for different values
of $e$. The local minima of ${\mathcal{M}}$ occur at pericenter ($f=0$) and at apocenter ($f=\pi$).
As noted by \citet{mut11} and \citet{gri15}, perturbers move 
supersonically, at any point of the orbit, as long as $X>2$ (see Fig. \ref{fig:Mach_f}).
In the remainder of the paper, we will focus on cases with $X>2$.

\begin{figure*}
\hskip -0.4cm
\includegraphics[width=197mm,height=70mm]{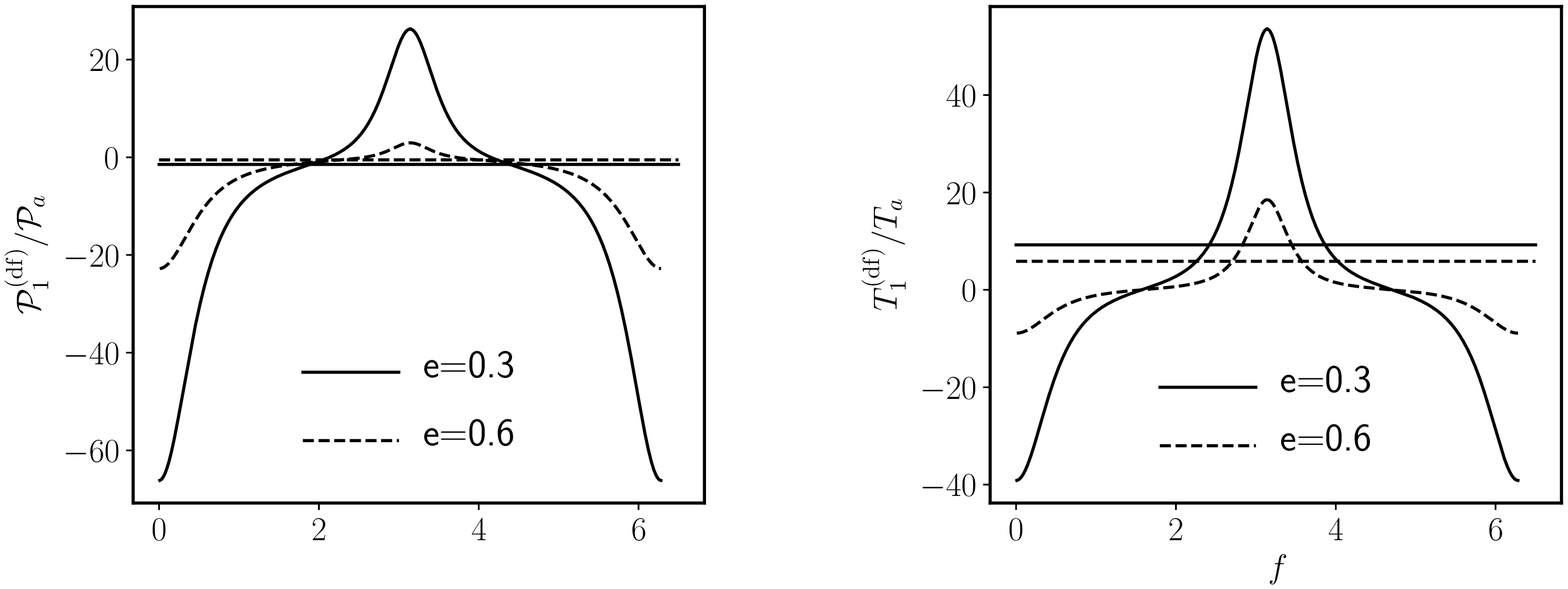}
  \caption{Theoretical power (left) and torque (right) versus $f$ in the local approximation,
 for $e=0.3$ (solid lines) and $e=0.6$ (dashed lines). We take $\alpha=0$. The horizontal lines indicate 
their mean values averaged over time.
 }
\vskip 0.75cm
\label{fig:start}
\end{figure*}

\section{The local approximation} 
\label{sec:local_approximation}
\subsection{Migration and eccentricity damping}
\label{sec:la_1}
Consider first a strictly 2D sheet of gas (i.e.~an infinitelly thin slab) with 
constant surface density $\Sigma_{0}$ and sound speed $c_{s}$. The slab, which is initially at rest, is perturbed
by a moving body which interacts only gravitationally with the gaseous medium through the potential
\begin{equation}
\Phi_{p}=-\frac{GM_{p}}{\sqrt{s^{2}+R_{\rm soft}^{2}}},
\end{equation}
where $s$ is the distance from the perturber.
We assume that the perturber travels in a rectilinear trajectory at constant {\it supersonic} velocity,
and denote by $\vecV_{\rm rel}$ the velocity of the gas relative to the perturber.  
Linear theory predicts that extended perturbers with softening radius $R_{\rm soft}$ and 
$V_{\rm rel}>c_{s}$ feel a dynamical friction force given by
\begin{equation}
\vecF_{\rm df}^{\scriptscriptstyle (2D)}\simeq \frac{\pi \Sigma_{0}G^{2}M_{p}^{2}}{R_{\rm soft} V_{\rm rel}^{3}}  \vecV_{\rm rel}
\label{eq:muto_Fdf_thin}
\end{equation}
(Muto et al. 2011). As long as the orbiter moves supersonically with respect to the gas and the
softening radius keeps constant along the trajectory,
$\vecF_{\rm df}$ does not depend on $c_{s}$.

\begin{table}
	\centering
	\caption{Parameters of our reference 2D runs} 
All the simulations in this Table use the fiducial vaues: $\alpha=0$, $h=0.04$ and $\nu=\nu_{0}=
10^{-5} \omega a^{2}$.
\vskip 0.3cm
\label{table:params} 
 \begin{tabular}{|l|c|c|c|c|c|}\hline
 Run &    $e$ & ${\mathcal{E}}$ & $R_{\rm in}$ &  $R_{\rm out}$ & zones per $R_{\rm soft}$   \\ 
                &        &                          &        &            &  $(N_{R}^{\rm peri}, N_{\phi})$      \\
               
\hline 
 1L       &    $0.1$  &  $0.6$  &  $0.4a$ & $3.5a$&  $(9,\, 6.5)$   \\
1S   &    $0.1$  &  $0.15$  &  $0.4a$ &  $3.5a$ & $(3,\, 2.5)$   \\ 
3La   &    $0.3$  &  $0.6$  &  $0.35a$ & $2.6a$ & $(14,\, 6)$   \\
3Lb   &    $0.3$  &  $0.6$  &  $0.23a$ & $3.9a$ & $(14,\, 6)$  \\
3Lc   &    $0.3$  &  $0.6$  &  $0.175a$ & $5.2a$ & $(14,\, 6)$   \\
3Ld   &    $0.3$  &  $0.6$  &  $0.35a$ & $5.2a$ & $(14,\, 6)$   \\
3S   &    $0.3$  &  $0.15$  &  $0.23a$ & $3.9a$ & $(2.5,\, 2)$    \\ 
6L   &    $0.6$  &  $0.6$  &  $0.12a$ &  $4.5a$  & $(12,\, 4)$  \\
6S   &    $0.6$  &  $0.15$  &  $0.175a$ &  $5a$ & $(2.5,\, 2)$   \\

 \hline 

\end{tabular}  
\label{table:parameters_sims1}
\vskip 1.0cm
\end{table}

Now consider a perturber embedded in the disk in a Keplerian orbit.
The local approximation consists in assuming that the interaction between a supersonic perturber
and the disk can be described at every point of the orbit
by Equation (\ref{eq:muto_Fdf_thin}) just taking the surface density, 
sound speed and $\vecV_{\rm rel}$ at the position of the perturber \citep[e.g.,][]{mut11,gri15}.

Once $\vecF_{\rm df}$ is known, we can evaluate the power 
${\mathcal{P}}^{\rm (df)}_{1}$ and the torque $T^{\rm (df)}_{1}$, predicted
in the local approximation, as a function of the true anomaly $f$.
Combining Eqs. (\ref{eq:vel_perturber}), (\ref{eq:vel_kepler}) and (\ref{eq:muto_Fdf_thin}), neglecting
the pressure term as it is of order of ${\mathcal{O}}(h^{2})$, and
using that $GM_{p}=q\omega^{2}a^{3}$,
we find that 
\begin{equation}
{\mathcal{P}}^{\rm (df)}_{1}=\vecv_{p}\cdot \vecF_{\rm df}^{\scriptscriptstyle (2D)}=\frac{\pi \eta q^{2} \omega^{3}a^{5}\Sigma_{p}}
{R_{\rm soft}}
\frac{-e^{2}\sin^{2} f +\xi \hat{\xi}}{[e^{2}\sin^{2}f +\hat{\xi}^{2}]^{3/2}},
\label{eq:power_DF}
\end{equation}
where $\Sigma_{p}$ is the unperturbed disk surface density at perturber's location, 
\begin{equation}
\xi(f) \equiv 1+e\cos f,
\end{equation}
 and
\begin{equation} 
\hat{\xi}(f)=\sqrt{\xi}-\xi.
\end{equation}
On the other hand, the torque is given by 
\begin{equation}
T^{\rm (df)}_{1}=\hat{\vece}_{z}\cdot (\vecr_{p}\times \vecF^{\scriptscriptstyle (2D)}_{\rm df})=
\frac{\pi  \eta^{4} q^{2}\omega^{2}a^{5} \Sigma_{p}}{R_{\rm soft}}
\frac{\hat{\xi}}{\xi (e^{2}\sin^{2} f +\hat{\xi}^{2})^{3/2}}.
\label{eq:torque_DF}
\end{equation}
In the most general case,  $R_{\rm soft}$ may depend on the position along the orbit.
If so, it should be evaluated at the instantaneous position of the perturber.
Once the power and the torque are known, the evolution of $a$ and $e$ can be computed using
Equations (\ref{eq:gauss1}) and (\ref{eq:gauss2}).

We warn that, instead of  ${\mathcal{P}}_{1}$, some authors provide the total power 
exerted by the accretion disk ${\mathcal{P}}_{\rm tot}={\mathcal{P}}_{d,0}+{\mathcal{P}}_{1}$,
where ${\mathcal{P}}_{d,0}$ is the power associated with the radial force created by the (axisymmetric)
unperturbed disk. 
As shown in the Appendix \ref{app:mestel}, ${\mathcal{P}}_{\rm tot}$ and ${\mathcal{P}}_{1}$
exhibit different dependences on $f$. There are cases where ${\mathcal{P}}_{\rm tot}$ may be dominated by
the contribution of ${\mathcal{P}}_{d,0}$. Nonetheless, ${\mathcal{P}}_{d,0}$ does not contribute to the change 
of the orbital elements (see Eqs. \ref{eq:gauss1} and \ref{eq:gauss2}) because its value averaged over one orbit is zero.

\begin{figure*}
\vspace{-1.6cm}
\includegraphics[width=188mm,height=250mm]{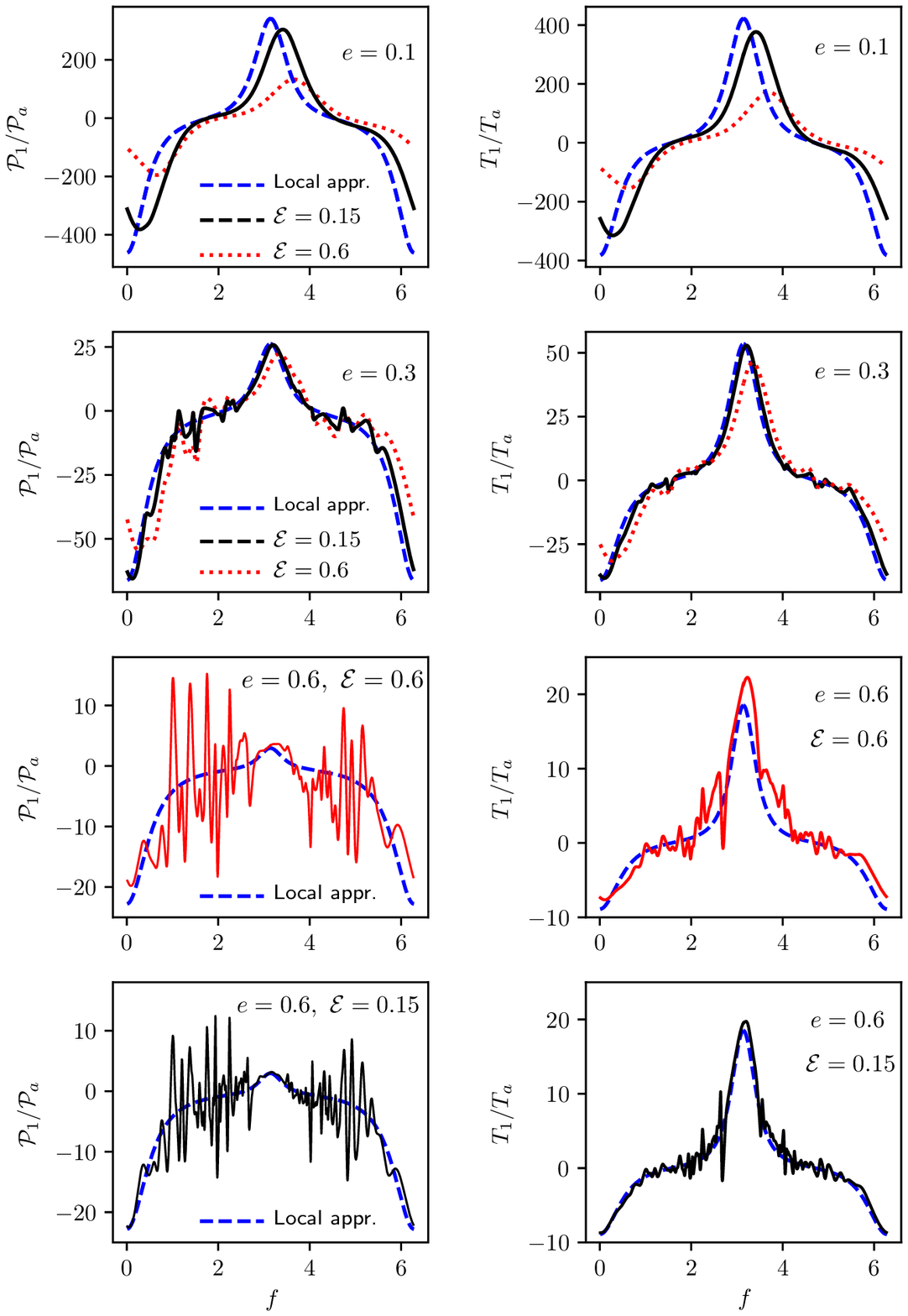}
  \caption{Dimensionless power (left columns) and torque (right columns) versus the true
anomaly on the 13th orbit, for different combinations of $e$ and $\mathcal{E}$. 
In all cases $\alpha=0$ and $h=0.04$. The value of $e$ is given
in the upper right corner on each panel. Black lines represent the values obtained from 2D simulations
with $\mathcal{E}=0.15$, while red lines correspond to $\mathcal{E}=0.6$. The predicted curves in
the local approximation are indicated by the dashed blue lines. For clarity, the case with $e=0.6$ has
been split in different panels.
 }
\vskip -0.05cm
\label{fig:pw_tq_A}
\end{figure*}

\begin{table}
	\centering
	\caption{Parameters of the complementary 2D runs} 
\vskip 0.3cm
\label{table:params} 
 \begin{tabular}{|l|c|c|c|c|c|c|c|}\hline
 Run &  $h$ & $\alpha$ & $e$ & ${\mathcal{E}}$ & $R_{\rm in}$ &  $R_{\rm out}$ & zones per $R_{\rm soft}$  \\ 
       &          &        &         &                 &        &            &  $(N_{R}^{\rm peri}, N_{\phi})$      \\
               
\hline 
A & $0.04$ & $0$ & $0.15$& $0.6$ & $0.23a$ & $3.9a$ & $(12,\, 4)$  \\
B & $0.04$ & $0.5$ & $0.6$& $0.6$ & $0.12a$ & $4.75a$ & $(7.5,\, 4)$  \\
C  &  $0.04$   & $1.5$  & $0.15$  &  $0.6$  &  $0.4a$ &  $3.5a$ & $(12, \,4)$   \\
D   &  $0.04$   &  $1.5$ & $0.3$  &  $0.6$  &  $0.23a$ &  $3.9a$ & $(9.5 ,\, 4)$   \\ 
E   &  $0.04$   & $1.5$ &  $0.6$  &  $0.15$  &  $0.12a$ &  $4.75a$ & $(2,\, 2)$  \\
F  & $0.1$ & $0$  &  $0.6$  &  $0.24$  &  $0.2a$ &  $5a$ & $(7.5,\, 4)$   \\
G  &  $0.1$ & $0$  &  $0.6$  &  $0.6$  &  $0.2a$ &  $5a$ & $(7.5,\, 4)$   \\

 \hline 

\end{tabular}  
\label{table:parameters_sims2}
\end{table}

\subsection{General considerations on the accuracy of the local approximation: open questions}
The local approximation implicitly assumes that the major contribution to the force
comes from material at distances $\ll R$ from the body. Therefore,
radial gradients in the unperturbed surface density and sound speed of the disk are disregarded
when calculating the structure of the wake. The local approximation also neglects the
differential rotation of the disk and thereby resonant effects. Thus, it also ignores that
for certain impact parameters, the streamlines are not supersonic relative to the perturber
even if ${\mathcal{M}}>1$ (see Appendix \ref{app:Mach1}). Finally, the local approximation
neglects the curvature of the wake and therefore it does not take into account that the 
perturber can catch its own wake.

A systematic study on the accuracy of the local approximation has not not been conducted
so far. Even in razor thin disks, the range of parameters within which the local
approximation is accurate has not been clearly established. One would expect that the local 
approximation overestimates the force because it ignores the curvature of the wake which
is expected to reduce the magnitude of $F_{1}$ \citep[e.g.,][]{kim07,san18}.
However, a rough comparison with the simulations in \citet{cre06}
indicates that the local approximation underestimates the torque by a factor of $2$ (see
fig. 8 in Muto et al. 2011). 

\citet{mut11} also noted that the behaviour of the power versus the orbital angle
reported in \cite{cre07} is very different to the predicted profile and this leads
them to conclude that the local approximation may result in an oversimplified model for $\vecF_{1}$.
However, \citet{mut11} compared ${\mathcal{P}}_{1}$ with ${\mathcal{P}}_{t}$, which are not the 
same quantity (see Appendix \ref{app:mestel}).

From the ongoing discussion, it is clear that a more fair comparison between simulations
and predictions is needed to evaluate the accuracy of the local approximation. This will be 
carried out in the next section.

\section{Numerical experiments}
\label{sec:experiments}
We have carried out a set of 2D simulations of a gaseous disk that is perturbed by a 
gravitational body using the code FARGO3D\footnote{FARGO3D is a publicly available code
at http://fargo.in2p3.fr.} \citep{ben16}
in polar coordinates centered on the central mass $M_{c}$.
The computational domain covers a ring with $R_{\rm in}\leq R \leq R_{\rm out}$
and $0\leq \phi \leq 2\pi$, where $R_{\rm in}$ and $R_{\rm out}$ are the inner and outer radii.
At both inner and outer boundaries, we use wave damping boundary conditions \citep{dev06}.
A locally isothermal equation of state is used, 
where the sound speed $c_{s}$ is a fixed function of radius; it is set out by requiring that
the disk aspect ratio $h$ defined as $c_{s}/(\Omega R)$ is constant with $R$. We also employ
a kinematic viscosity $\nu$ that is constant over the entire disk. In most of the models, 
$\nu=10^{-5}\omega a^{2}$. The unperturbed surface density of the disk follows a power law
$\Sigma_{0} =\Sigma_{a}(R/a)^{-\alpha}$.

We consider a perturber in a fixed elliptical orbit with
eccentricity $e$. The perturber's gravitational potential is smoothed over a fraction $\mathcal{E}$
of the local value of $H$ (defined as $c_{s}/\Omega$), so that $\mathcal{E}\equiv R_{\rm soft}/H$ is constant
along the orbit. No removal of mass near the perturber was implemented. 

Our assumption that $\mathcal{E}$ is constant along the orbit is physically justified for 
perturbers moving in circular orbits \citep[e.g.,][]{mas02,mul12}, but this is not the case here.
For elliptical orbits, one may consider to use a different dependence of $\mathcal{E}$ with
the position and velocity of the perturber. 
Since the local approximation does not require any particular choice for $\mathcal{E}$,
we will use this simplest assumption for the sake of concreteness.

We have performed calculations with different $\alpha$, $h$, $e$ and ${\mathcal{E}}$.
The parameters of our fiducial models (i.e. those models with $\alpha=0$ and $h=0.04$)
are compiled in Table \ref{table:parameters_sims1}.  In these simulations, we vary only two parameters: 
the eccentricity between $0.1$ and $0.6$, and ${\mathcal{E}}$ between $0.15$ and $0.6$. 
We thus employ a mnemonic nomenclature for the runs using the number $e/0.1$, followed 
by S or L, indicating whether ${\mathcal{E}}$ is small (${\mathcal{E}}=0.15$) or
large (${\mathcal{E}}=0.6$). For instance, Run 3L indicates that $e=0.3$ and ${\mathcal{E}}=0.6$.
Other complementary models with different $\alpha$ or $h$ are listed in Table \ref{table:parameters_sims2}.

The value of the mass ratio $q$ was taken small enough so that the interaction is linear but
not too small that the results could be affected by numerical noise.
As a compromise, we adopted $q=10^{-5}$ in all simulations except Run 1S for
which we took $q=2.5\times 10^{-6}$.

In all simulations, the number of zones per $R_{\rm soft}$ 
in the radial $N_{R}$ and azimuthal $N_{\phi}$ directions
is at least $2$, at any point of the orbit. Since the zones are linearly spaced in $R$ and $\phi$,
$N_{\phi}$ is independent of $R$, but $N_{R}$ varies with $R$, being lowest at pericenter with 
a value given in Tables \ref{table:parameters_sims1} and \ref{table:parameters_sims2}.

Our aim is to compute ${\mathcal{P}}_{1}$ and $T_{1}$ in the simulations and compare them to the values 
${\mathcal{P}}_{1}^{\rm (df)}$ and $T_{1}^{\rm (df)}$ derived in the local approximation. 
More specifically, the power and the torque were obtained from the simulations using
${\mathcal{P}}_{1}=\vecv_{p}\cdot \vecF_{1}$ and $T_{1}=\hat{\vece}_{z}\cdot (\vecr_{p}\times \vecF_{1})$ with
\begin{equation}
\vecF_{1}=\int (\Sigma-\Sigma_{0}) \vecnabla \Phi_{p} \, dA,
\end{equation}
where $dA$ is the surface element. We recall that $\Sigma_{0}$ is the unperturbed, i.e. the initial,
surface density of the disk.

By using that $R_{\rm soft}=\mathcal{E} h R$ in our disk models,
Equations (\ref{eq:power_DF}) and (\ref{eq:torque_DF}) for the power and the torque can be written as
\begin{equation}
{\mathcal{P}}^{\rm (df)}_{1}= {\mathcal{P}}_{a} \xi^{1+\alpha} 
\frac{(-e^{2}\sin^{2} f +\xi \hat{\xi})}{(e^{2}\sin^{2} f +\hat{\xi}^{2})^{3/2}},
\label{eq:power_DF_model}
\end{equation}
\begin{equation}
T^{\rm (df)}_{1}= 
\frac{\xi^{\alpha}\hat{\xi}T_{a} }{(e^{2}\sin^{2} f +\hat{\xi}^{2})^{3/2}},
\label{eq:torque_DF_model}
\end{equation}
where 
\begin{equation}
{\mathcal{P}}_{a}= \frac{\pi  q^{2} \omega^{3} a^{4} \Sigma_{a}}{\eta^{1+2\alpha} \mathcal{E} h},
\end{equation}
and
\begin{equation}
T_{a} = \frac{\pi \eta^{2(1-\alpha)} q^{2} \omega^{2} a^{4}\Sigma_{a}}{\mathcal{E} h}.
\label{eq:amplitude_torque}
\end{equation}
The dimensionless power ${\mathcal{P}}^{\rm (df)}_{1}/\mathcal{P}_{a}$ and torque
$T_{1}^{\rm (df)}/T_{a}$ only depend on $\alpha$, $e$ and the orbital phase $f$. 
For illustration, Figure \ref{fig:start} shows ${\mathcal{P}}_{1}^{\rm (df)}$
and $T_{1}^{\rm (df)}$ as a function of $f$ for $\alpha=0$ and two values of $e$ ($0.3$ and $0.6$). 
Both ${\mathcal{P}}_{1}^{\rm (df)}$ and $T_{1}^{\rm (df)}$ are positive at apocenter ($f=\pi$)
and negative at pericenter ($f=0$). This is because the gas rotates faster than the 
perturber at apocenter and pushes it \citep{cre07,mut11}.
At pericenter, on the contrary, the perturber experiences a drag because it moves
at a speed greater than the gas. We see that the
mean values of the power over one orbit are small compared to their dynamical
range. In the next section (\S \ref{sec:short_term}), we examine whether the local approximation can
account for the changes of ${\mathcal{P}}_{1}$ and $T_{1}$ along the orbit. Later, in \S \ref{sec:long_term},
we check if the mean values over one orbit are consistent with the estimates in
the framework of the local approximation.

\begin{figure*}
\vspace{0.0cm}
\includegraphics[width=197mm,height=70mm]{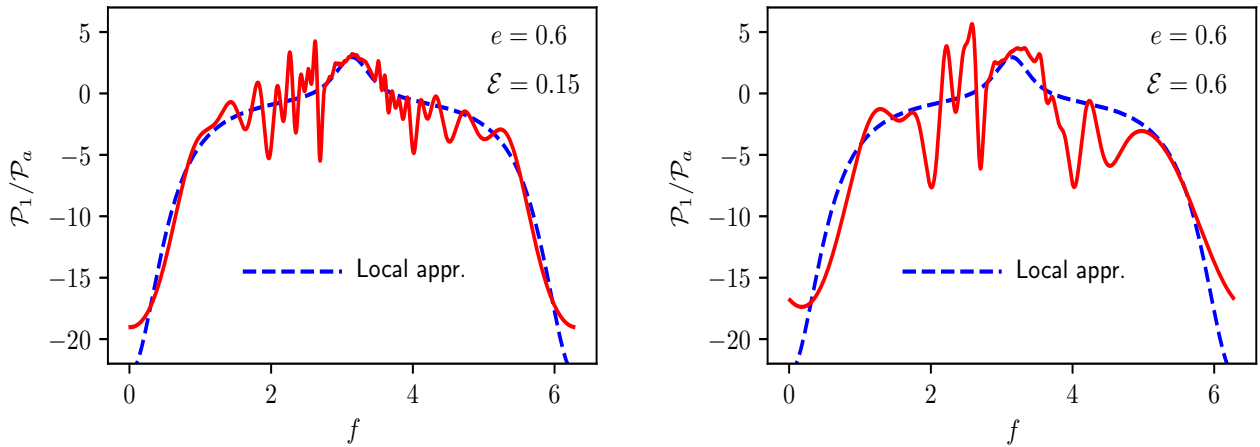}
  \caption{Dimensionless power during the $13$th orbit in Run 6S (left) and Run 6L (right), after
filtering out the high-frequency oscillations (solid lines). The dashed lines indicate the power
estimated in the local approximation.   
 }
\vskip -0.00cm
\label{fig:power_filtered}
\end{figure*}

\begin{figure}
\includegraphics[width=90mm,height=70mm]{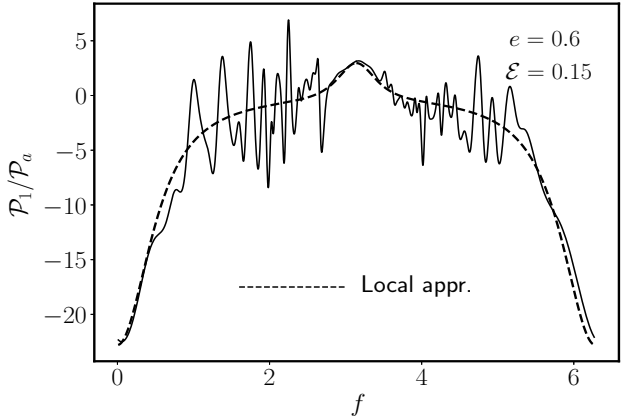}
  \caption{Dimensionless power as a function of $f$, during the $13$th orbit, for a
simulation with same parameters as Run 6S expect the viscosity, which is factor of $5$ larger. 
The dashed line indicates the theoretical values in the local approximation.
 }
\vskip 0.25cm
\label{fig:ecc06_nu5}
\end{figure}

\begin{figure*}
\vspace{0.0cm}
\includegraphics[width=188mm,height=60mm]{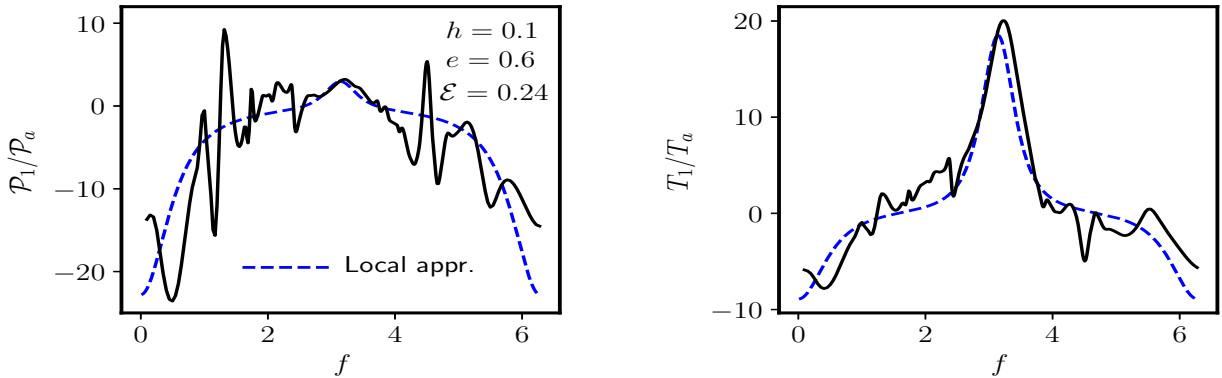}
  \caption{Dimensionless power (left panel) and torque (right panel) during the $13$th orbit 
in Run F (solid lines). The dashed lines indicate the 
theoretical curves in the framework of the local approximation.  
 }
\vskip -0.00cm
\label{fig:pw_tq_h01}
\end{figure*}

\begin{figure*}
\vspace{0.0cm}
\includegraphics[width=188mm,height=60mm]{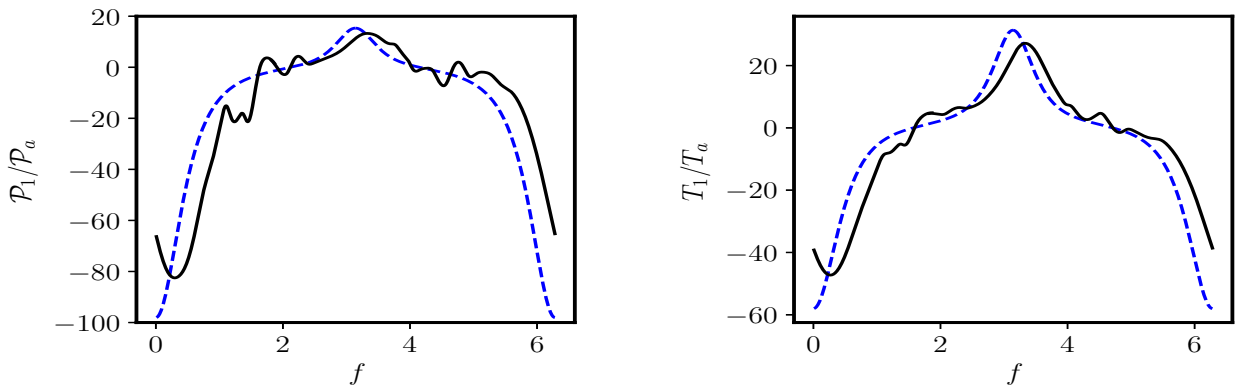}
  \caption{Same as Figure \ref{fig:pw_tq_h01} but for Run D.
 }
\vskip -0.00cm
\label{fig:pw_tq_alpha15}
\end{figure*}

\begin{figure}
\hskip -0.4cm
\includegraphics[width=99mm,height=125mm]{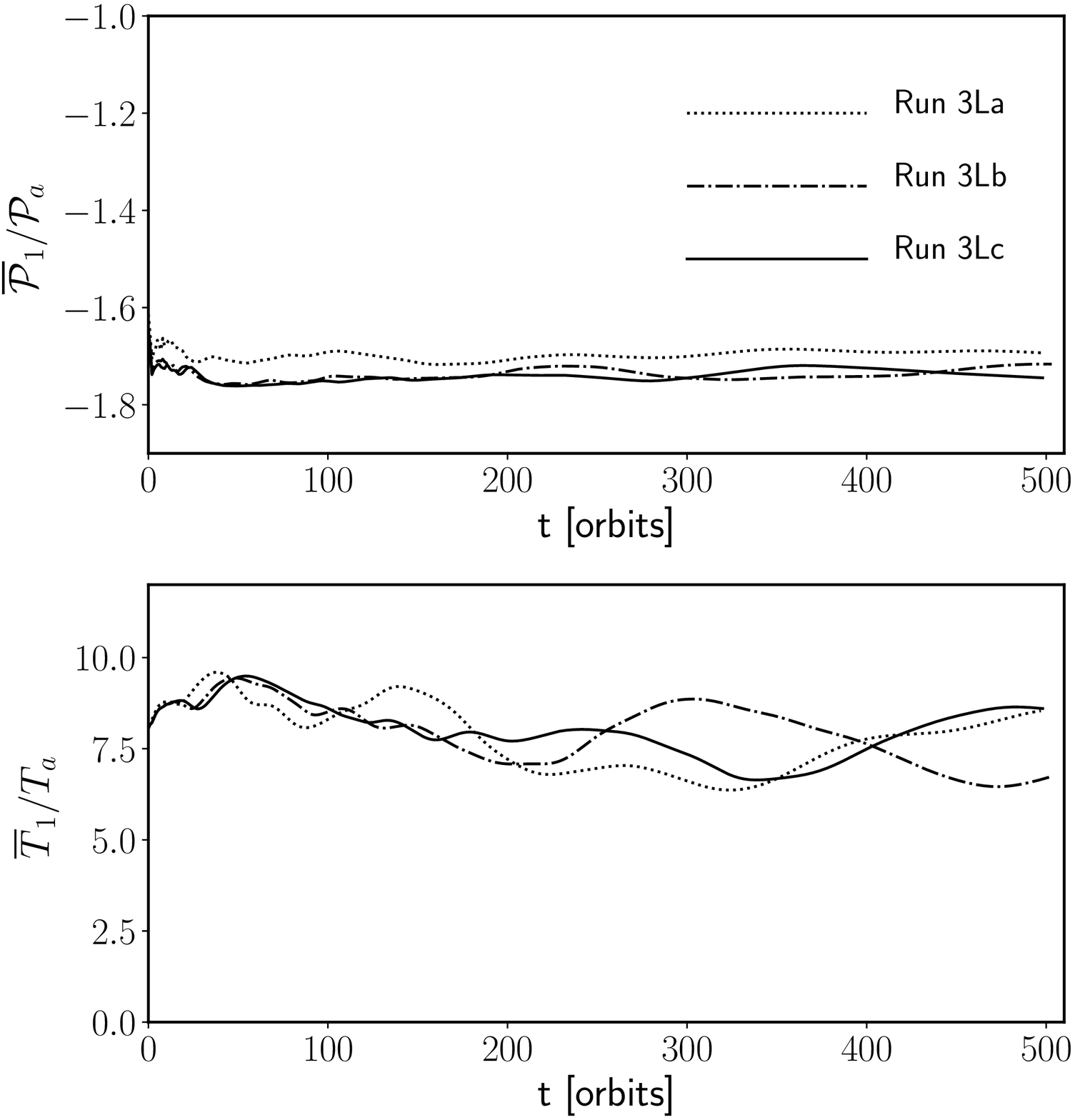}
  \caption{Time evolution of $\overline{\mathcal{P}}_{1}$ and $\overline{T}_{1}$, in dimensionless units, 
for various sizes of the computational domain (see Table \ref{table:parameters_sims1}).
The parameters of the disk, the eccentricity and the softening
radius are the same in all cases ($h=0.04$, $e=0.3$, and 
${\mathcal{E}}=0.6$).
 }
\vskip 0.75cm

\label{fig:diff_sizes}
\end{figure}

\subsection{Dependence of ${\mathcal{P}}_{1}$ and $T_{1}$ on the orbital phase}
\label{sec:short_term}

For clarity, we will first focus on the simulations of a disk with $\alpha=0$ 
(i.e.~constant surface density at $t=0$) and $h=0.04$. Figure \ref{fig:pw_tq_A} shows 
${\mathcal{P}}_{1}$ and $T_{1}$ versus $f$, for
$e=0.1, 0.3$ and $0.6$, which correspond to $X=2.5, 7.5$ and $15$, respectively.
The curves in Figure \ref{fig:pw_tq_A} were extracted when the perturber was 
completing the $13$th orbit.

\begin{figure*}
\vspace{-0.5cm}
\includegraphics[width=188mm,height=220mm]{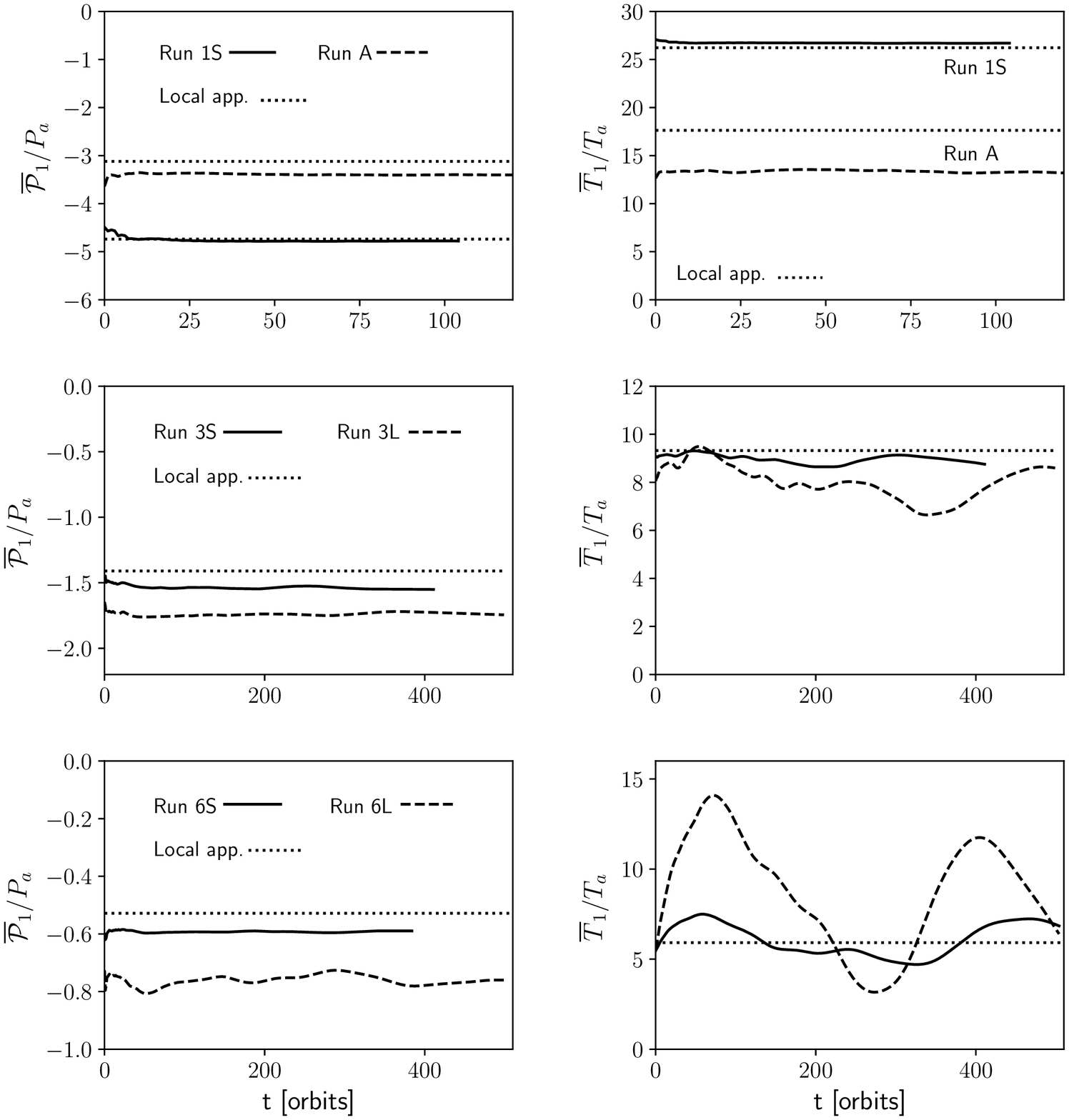}
  \caption{
Time evolution of $\overline{{\mathcal{P}}}_{1}$ (left column) and $\overline{T}_{1}$ (right column)
for different combinations of $e$ and ${\mathcal{E}}$. The solid curves are for models with 
${\mathcal{E}}=0.15$ and the dashed curves for ${\mathcal{E}}=0.6$. The horizontal dotted 
lines correspond to the predicted values in the local approximation. In all cases $\alpha=0$ and $h=0.04$.
}
\vskip -0.25cm
\label{fig:mean_pw_tq}
\end{figure*}

In general, the differences between predictions and numerical results disminish as $\mathcal{E}$ 
decreases. The reason is simple; the
relative contribution of the field in the vicinity of the body increases as $\mathcal{E}$ decreases. Therefore, the
contribution of the far field, which is not captured well in the local approximation, becomes gradually
less important relative to the contribution of the near field as $\mathcal{E}$ decreases. 

For $e=0.1$ (i.e.~$X=2.5$), the local approximation can reproduce neither
the magnitude of the power nor the torque if $\mathcal{E}=0.6$. 
This is expected because the Mach $1$ distance is $\lesssim R_{\rm soft}$ (see Appendix \ref{app:Mach1}).
We also see that the curves of ${\mathcal{P}}_{1}$ and $T_{1}$ are shifted with
respect to the predicted curves for $\mathcal{E}=0.6$. 
If ${\mathcal{E}}$ is reduced a factor of $4$ (${\mathcal{E}}=0.15$), 
the curves match quite well each other if the predicted curves are shifted 
right by $\Delta f=0.33$. This shift has little effect when computing migration and eccentricity
damping timescales because the averaged values over one orbit are preserved.

For $e=0.3$ (i.e.~$X=7.5$), the local approximation predicts correctly ${\mathcal{P}}_{1}$ and 
$T_{1}$ for $\mathcal{E}=0.15$. Even for $\mathcal{E}=0.6$, the shapes of ${\mathcal{P}}_{1}$ 
and $T_{1}$ are captured well in the local approximation.  At apocenter, the power and the torque 
are a bit lower than predicted. They also slightly deviate at pericenter.

For $e=0.6$ (i.e.~$X=15$), $T_{1}$ but especially $\mathcal{P}_{1}$ exhibit spikes that are
produced when the perturber
crosses shock fronts and density substructures. These spikes are well-resolved in both strength and
time, but they obscure the averaged value over a longer timescale. In order to make
a better comparison with the values predicted by the local approximation, we use a time Fourier 
filter to remove high-frequency modes. Figure \ref{fig:power_filtered} shows
that the filtered power for $e=0.6$ behaves in the manner predicted by the local approximation,
even if ${\mathcal{E}}=0.6$.

A larger viscosity may smear the gradients in the velocity and may contribute to
smooth the power and torque. Figure \ref{fig:ecc06_nu5} shows the non-filtered power in 
a simulation similar to Run 6S except the viscosity was increased by a factor of $5$. 
The amplitude of the spikes reduces by a factor of $2$.

In order to illustrate the influence of the temperature of the disk on the abundance and
amplitude of spikes,
Figure \ref{fig:pw_tq_h01} shows the power and the torque also for $e=0.6$, but $h=0.1$, 
implying $X=6$ (Run F in Table \ref{table:parameters_sims2}). 
This simulation has the same ${\mathcal{E}}h=0.024$ as Run 6L, and thereby
they have the same softening radius. The level of substructure in ${\mathcal{P}}_{1}$
and $T_{1}$ is reduced as compared to Run 6L. A slight asymmetry with respect to 
$f=\pi$ is visible in both the power and the torque. The main discrepancy between simulations 
and the predicted values occurs for the power when the perturber is passing close to pericenter.

Finally, we have verified that the local approximation also predicts sucessfully the shape of
${\mathcal{P}}_{1}$ and $T_{1}$ for a disk with $\alpha=1.5$. As an example,
Figure \ref{fig:pw_tq_alpha15} shows the power and the torque for $e=0.3$ and $\alpha=1.5$ (Run D in 
Table \ref{table:parameters_sims2}).

In summary, we find that for ${\mathcal{E}}<0.6$,
the local approximation reproduces qualitatively the dependence of ${\mathcal{P}}_{1}$ and $T_{1}$ 
with the orbital phase, after several orbits, provided that $X>3.75$. 
For values $X\simeq 2.5$, we need smaller values for ${\mathcal{E}}$. For $X\simeq 15$,
the power presents remarkable spikes but still the local approximation can explain the
underlying shape. In the next section, we carry out an analysis of the orbit averaged values of the power
and the torque, and consider a longer timescale.

\subsection{Averaged values of the power and torque over one orbital period: Long-term evolution}
\label{sec:long_term}
The relevant quantities to compute the orbital evolution of the perturbing object are 
$\overline{{\mathcal{P}}}_{1}$ 
and $\overline{T}_{1}$, where the over-bar indicates the average value over intervals of one orbital period.
In analytical calculations, it is frequent to assume that all the quantities of the fluid
are periodic with frequency $\omega$, i.e. the perturbation in the gas
is the same in succesive passes of the body at the same position. Under this assumption,
$\overline{{\mathcal{P}}}_{1}$ and $\overline{T}_{1}$ are independent of time.

For $h=0.04$, we find that $\overline{\mathcal{P}}_{1}$ and $\overline{T}_{1}$ maintain approximately
constant along the simulation if $e<0.2$ (i.e.~$X<5$). In general, however, they are not 
constant but display long-term
variations. Such temporal changes may be genuine or a consequence of spurious boundary effects.
In order to assess the effect of the limited size of the computational box, Figure \ref{fig:diff_sizes} 
shows the time evolution of $\overline{\mathcal{P}}_{1}$ and $\overline{T}_{1}$ for our fiducial parameters 
($\alpha=0$, $h=0.04$) with
$e=0.3$, and ${\mathcal{E}}=0.6$ for various sizes of the domain, keeping 
the same resolution (Runs 3La, 3Lb and 3Lc in Table \ref{table:parameters_sims1}).
We will focus on the behaviour of the torque because the differences in the power are ignorable.
At $t<300$ orbits, the magnitude of the variations in the torque is least in the simulation
with the largest radial extension (Run 3Lc); $\overline{T}_{1}/T_{a}$ varies gradually between $9.5$ at $40$ orbits
to $8$ at $t=275$ orbits (a change of $16\%$). After $275$ orbits, the dimensionless torque
in the three simulations oscillates between $6$ and $9$. Although we cannot rule out that, 
beyond $275$ orbits, part of the temporal variation of the torque is caused by 
boundary artifacts even in Run 3Lc, the mean value of $\overline{T}_{1}$ over the whole runtime is 
rather similar in the three simulations. 

In Runs 3La, 3Lb and 3Lc, both $R_{\rm in}$ and $R_{\rm out}$ were varied. However, we have carried
out simulations with the same $R_{\rm out}$, but with different $R_{\rm in}$ (from $0.17a$ to $0.35a$)
and found that the oscillations in the torque are not very sensitive to $R_{\rm in}$ for values within that range.
We have also found that the results are robust to reasonable changes
in the size of the wave killing region in our damping conditions.

Figure \ref{fig:mean_pw_tq} shows the temporal evolution of $\overline{\mathcal{P}}_{1}$ and 
$\overline{T}_{1}$ for $\alpha=0$, $h=0.04$ and different combinations of $e$ and ${\mathcal{E}}$. 
The horizontal lines correspond to the values predicted in the local approximation. 
The first result is that the agreement between simulations and theoretical estimates
is reasonably good in all the cases when ${\mathcal{E}}=0.15$. In addition, for this
value of ${\mathcal{E}}$, $\overline{T}_{1}$ is fairly constant over time for $e\leq 0.3$.
For $e=0.6$, $\overline{T}_{1}$ varies around a value close to that predicted by the local
approximation with a moderate amplitude.

\begin{figure}
\vspace{0.0cm}
\includegraphics[width=92mm,height=60mm]{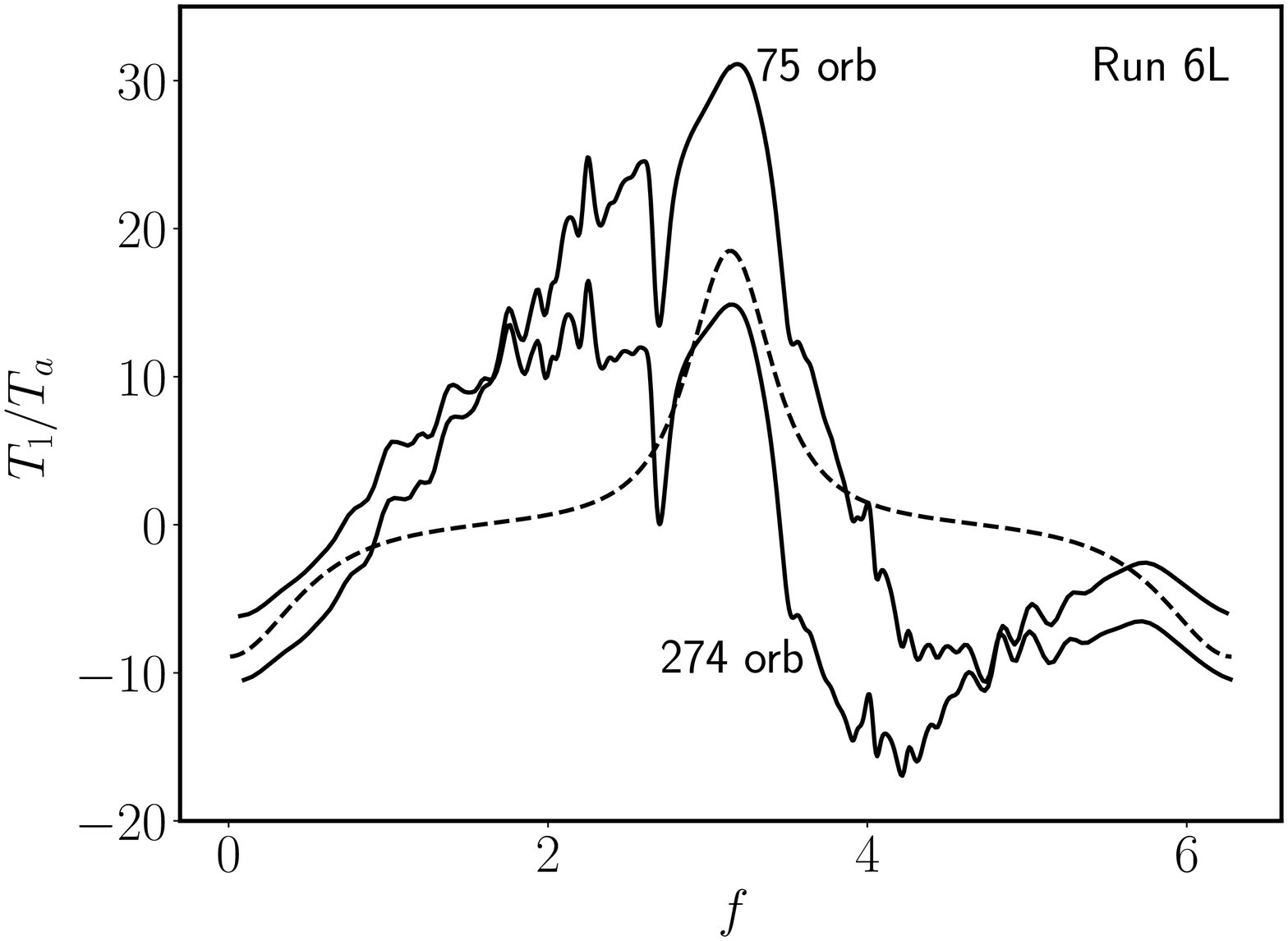}
  \caption{Dimensionless torque over the $75$th orbit (upper solid curve), and the $274$th orbit (lower solid
curve) for Run 6L ($e=0.6$ and
${\mathcal{E}}=0.6$). At these orbits, $\overline{T}_{1}$ presents a local maximum and minimum, respectively.
For reference, the local approximation curve is also given (dashed line).
 }
\vskip -0.00cm
\label{fig:torque_two_times}
\end{figure}

For ${\mathcal{E}}=0.6$, $\overline{T}_{1}$ exhibits long-term variations of large amplitude 
if $e=0.6$ (Run 6L). These variations occur in a characteristic timescale of 
$\tau_{\rm var}\simeq 200$ orbits.
The fact that the torque increases by a factor of $2.4$ in the first $75$ orbits suggests that the changes
in $\overline{T}_{1}$ have a physical origin rather than being a numerical artifact.
Figure \ref{fig:torque_two_times} shows the torque
as a function of the orbital phase during the $75$th and $274$th orbits,
i.e. when the torque reaches a local maximum and a local minimum, respectively.  
The curves $T_{1}$ vs $f$ are now clearly asymmetric with respect to $f=\pi$;
the torque when the perturber travels from pericenter to apocenter is different to
when it goes from apocenter to pericenter. The variations in the torque are a consequence
of the complexity of the far-field flow, which takes hundreds of orbits to achieve a periodic
configuration for $e=0.6$.

We have run the same simulation (Run 6L) with viscosities between $0.2\times 10^{-5}$ and
$5\times 10^{-5}$ (in units of $\omega a^{2}$) and found only a $10\%$ change in
$\overline{T}_{1}$ after $300$ orbits.
This is expected because the origin of the long-term fluctuations in the torque is related to the 
large-scale perturbations in the flow, which are unaffected by viscosity.

We have also computed the torque in simulations where the orbit is not fixed to be elliptical,
but forms a rosette figure after including the potential associated with the unperturbed
disk. In these simulations, the changes of $\overline{T}_{1}$ over time are similar.

\begin{figure}
\hskip -0.4cm
\includegraphics[width=99mm,height=115mm]{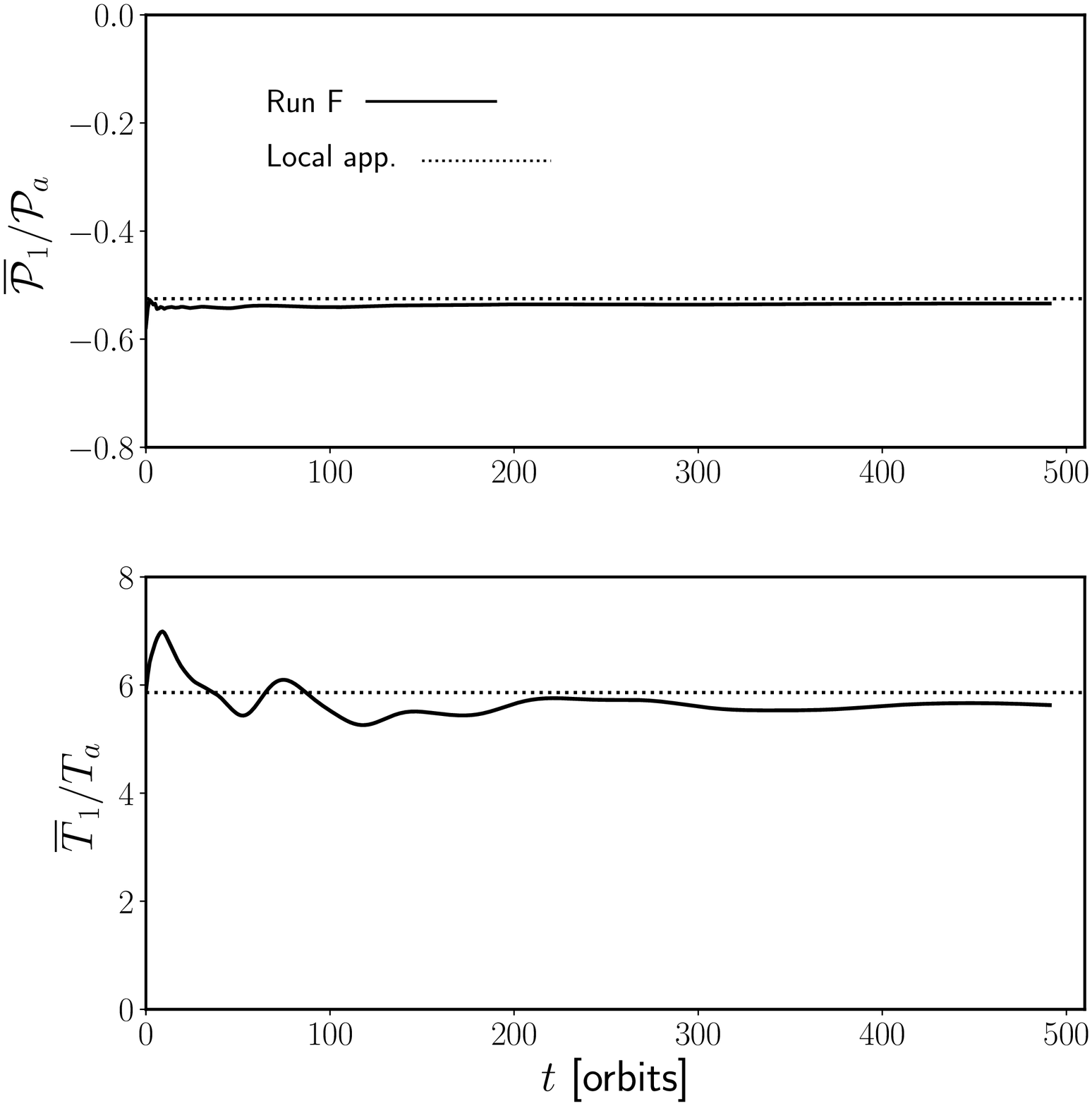}
  \caption{Time evolution of $\overline{\mathcal{P}}_{1}$ (top panel) and $\overline{T}_{1}$ (bottom panel),
in dimensionless units,
for model F. The dotted lines indicate the theoretical values in the local approximation.
 }
\label{fig:h01}
\end{figure}

The amplitude of the temporal variations in $\overline{T}_{1}$ depend largely on $\alpha$ and $h$. 
In disks with larger values of $h$, the sound speed is larger and the amplitude
of density perturbations in the disk are smeared out in a shorter timescale.
For a model with $h=0.1$ and $e=0.6$ (Run F), 
$\overline{T}_{1}$ is essentially constant after $130$ orbits (see Figure \ref{fig:h01}). 

\begin{figure}
\hskip -0.4cm
\includegraphics[width=99mm,height=155mm]{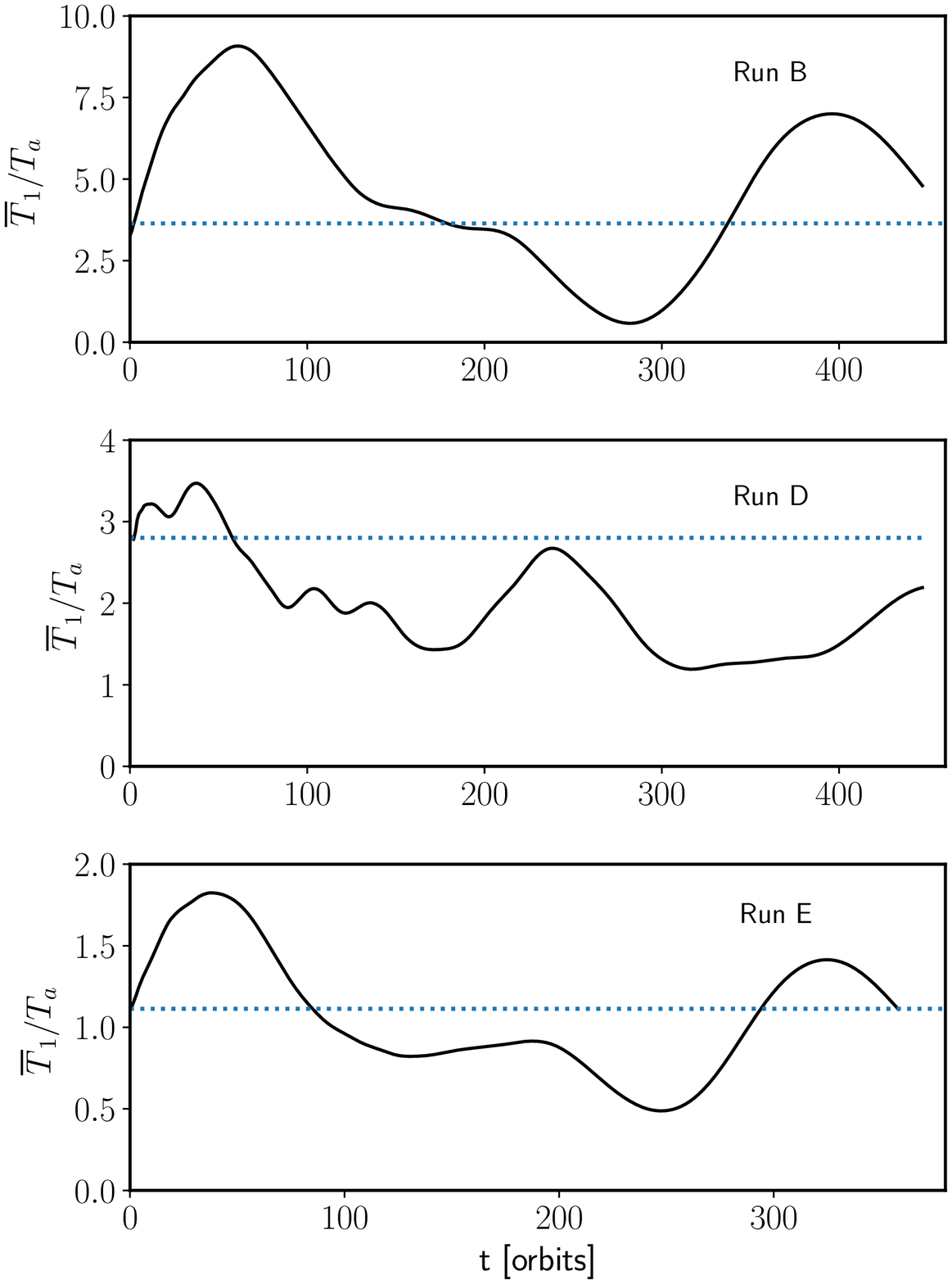}
  \caption{Time evolution of $\overline{T}_{1}/T_{a}$
for models with $\alpha=0.5$ (top panel) and
$\alpha=1.5$ (middle and bottom panels). The horizontal dotted lines show the predicted values adopting
the local approximation.
 }
\vskip 0.75cm

\label{fig:alpha05to15}
\end{figure}

Figure \ref{fig:alpha05to15} shows the time evolution of $\overline{T}_{1}$ 
for two different values of $\alpha$  ($\alpha=0.5$ and $1.5$). 
Greater is the value of $\alpha$, higher is the amplitude of the variations in the torque.
We warn that in the simulations with $\alpha=1.5$, our damping boundary conditions do not 
preserve mass over the runtime.
For instance, in Run E, the mass contained within the apocenter radius increases by $22\%$ after 
$360$ orbits. A more delicate comparison should take into account this secular mass enhancement.

It is remarkable and worthwhile noting that at $t\lesssim 12$ orbits, the values of $\overline{T}_{1}$ in the 
simulations are fully consistent with those obtained in the framework of the local approximation, even 
in runs with ${\mathcal{E}}=0.6$ (see Figs. \ref{fig:mean_pw_tq}, \ref{fig:h01} and \ref{fig:alpha05to15}).

\subsection{The maximum softening radius}

From our simulations, we can compute the mean value of the torque over the runtime $t_{\rm run}$ as 
\begin{equation}
\left< \overline{T}_{1}\right> = \frac{1}{t_{\rm run}} \int_{0}^{t_{\rm run}} \overline{T}_{1} dt.
\end{equation}
For models with large temporal variations in the torque, $\left< \overline{T}_{1}\right>$ 
is meaningful only if $t_{\rm run}$, $\tau_{a}$ and $\tau_{e}$ are $\gg \tau_{\rm var}$.
Otherwise, one should consider the detailed temporal evolution of the power and the torque
to find the evolution of the orbital parameters of the embedded object.
Since $\tau_{e}\ll \tau_{a}$,
the required condition is $\tau_{e}\gg \tau_{\rm var}\simeq 150$ orbits. Given that
$\tau_{e}$ increases as $q$ decreases, this condition provides an upper limit value for $q$.

As we have seen in the previous section (\S \ref{sec:long_term}),
the amplitude of the variations in the power and the torque decreases as ${\mathcal{E}}$
decreases. In fact, for ${\mathcal{E}}$ small enough, $\overline{\mathcal{P}}_{1}$ and $\overline{T}_{1}$ converge
to the values predicted in the local approximation and, in addition,
the rms of $\overline{\mathcal{P}}_{1}$ and $\overline{T}_{1}$ also decrease. Consequently, given
the disk parameters $\alpha$ and $h$, and the orbital eccentricity, there
exists a maximum value of $\mathcal{E}$, denoted by $\mathcal{E}_{\rm max}^{(2D)}$, such that if
$\mathcal{E}<\mathcal{E}_{\rm max}^{(2D)}$ then (1) the local approximation provides the mean
power and torque with an error less than $20\%$, and (2) the rms value of $\overline{T}_{1}$
is less than $0.15 \left< \overline{T}_{1}\right>$. If conditions (1) and (2) are met, the local
approximation shall be deemed satisfactory.
In Table \ref{table:parameters_epsilon}, we provide the
values of $\mathcal{E}_{\rm max}^{(2D)}$ for different disk parameters and orbital eccentricities.

\begin{table}
	\centering
	\caption{${\mathcal{E}}_{\rm max}^{(2D)}$ and ${\mathcal{E}}_{\rm max}^{(3D)}$ for some
disk parameters and eccentricities} 

\vskip 0.3cm
\label{table:params} 
 \begin{tabular}{|c|c|c|c|c|}\hline
    $h$ & $\alpha$ & $e$ &  ${\mathcal{E}}_{\rm max}^{(2D)}$ &    ${\mathcal{E}}_{\rm max}^{(3D)}$\\

\hline 
     $0.04$  &  $0$  &  $0.1$ & $0.25$&  $0.12$   \\
  $0.04$  &  $0$  &  $0.15$ &  $0.45$ & $0.34$   \\  
    $0.04$  &  $0$  &  $0.3$ & $0.6$ & $0.46$   \\
  $0.04$  &  $0$  &  $0.6$ & $0.15$ & $ 0.078$  \\
 $0.04$  &  $0.5$  &  $0.6$ & $0.1$ & $0.01 $   \\
   $0.04$  &  $1.5$  &  $0.3$ & $0.24$ & $0.12$    \\ 
 $0.04$  &  $1.5$  &  $0.6$ & $0.06$ & $8\times 10^{-4}$    \\
 $0.1$  &  $0$  &  $0.6$ & $0.7$ & $0.6$    \\

 \hline 

\end{tabular}  
\label{table:parameters_epsilon}
\vskip 1.0cm
\end{table}

Along this section, we have implicitly assumed that the accretion radius of the perturber
$R_{\rm acc}\equiv G M_{p}/V_{\rm rel}^{2}$
is smaller than $R_{\rm soft}$ so that the perturbation is linear at any position, even at the
vicinity of the perturber. In the case that $R_{\rm soft}<R_{\rm acc}$ then the relevant
radius is not longer $R_{\rm soft}$ but $R_{\rm acc}$ \citep[e.g.,][]{ber13} and thereby  
the condition for the local approximation to be valid is $R_{\rm acc} < {\mathcal{E}}_{\rm max}^{(2D)} H$.

All the above considerations were depicted for a softened perturber embedded in
a razor-thin disk. In the next Section, we extend the analysis of the applicability of 
the local approximation to a more realistic 3D disk and also to accreting perturbers.

\section{Local approximation in 3D disks}
\label{sec:3Dapprox}
The extension of the drag force, $F_{\rm df}^{(3D)}$, to a plane-parallel slab with
finite thickness was derived in \citet{can13}. They assume that the 
perturber moves in rectilinear trajectory in the midplane of a vertically
stratified slab with density 
$\rho (z)=\rho_{0}\exp (-z^{2}/2H^{2})$. For a nonaccreting perturber with
softening radius much smaller than $H$, they infer that 
\begin{equation}
F^{\scriptscriptstyle (3D)}_{\rm df}= \frac{\sqrt{8\pi}\Sigma (GM_{p})^{2}}{V^{2}_{\rm rel} H} \ln\left(\frac{1.32H}{R_{\rm soft}}\right)
\label{eq:canto_softened}
\end{equation}
\citep[see][for details]{san18}.

Since numerical simulations of a perturber in eccentric orbit embedded in a 3D disk
are computationally expensive, it is useful to derive under which conditions
the local approximation, using $F_{\rm df}^{(3D)}$, is appropriate to describe the 
interaction between a perturber and a 3D disk.

In \S \ref{sec:short_term} and \ref{sec:long_term}, we found that if 
$R_{\rm soft}\leq \Rtilde_{\rm soft}^{(2D)}\equiv {\mathcal{E}}_{\rm max}^{(2D)} H$, the local
approximation in a 2D disk is reasonably accurate because the near wake region of the perturber, defined
as the region in the vicinity of the perturber that is not affected by curvature terms, 
contributes to $80\%$ of the drag force or more. The near wake region in a 2D slab
has a size $\simeq 5\Rtilde_{\rm soft}^{(2D)}$.

\begin{figure}
\hskip -0.1cm
\includegraphics[width=93mm,height=130mm]{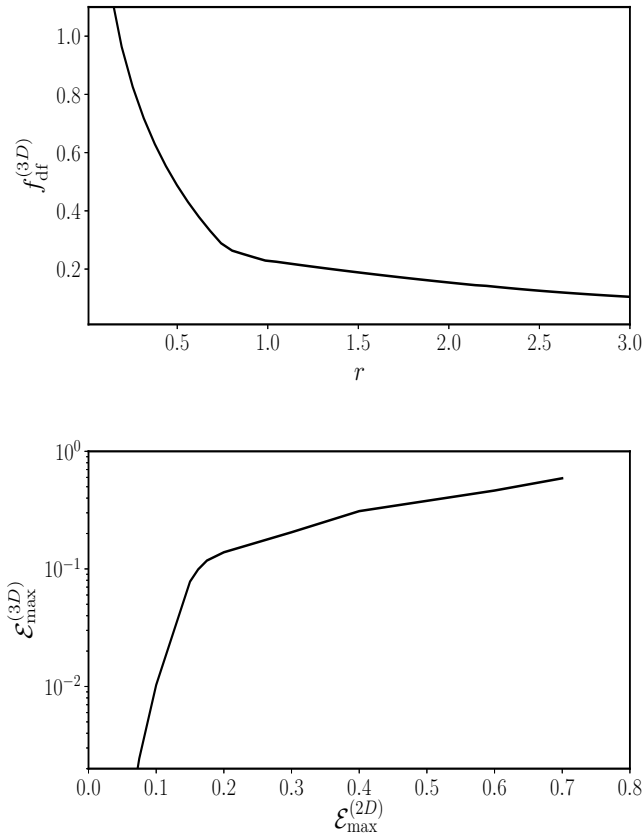}
  \caption{
Drag force on a perturber embedded in a vertically-stratified plane-parallel medium, arising
from the enhanced-density wake at distances greater than $r$ from the perturber,
in arbitrary units (top panel). Relationship between ${\mathcal{E}}_{\rm max}^{(2D)}$ 
and ${\mathcal{E}}_{\rm max}^{(3D)}$ (bottom panel).
 }
\label{fig:2Dto3D}
\end{figure}

In a disk with finite thickness, we also expect that the local approximation should be valid
for sufficiently small perturbers, say $R_{\rm soft}\leq \Rtilde_{\rm soft}^{(3D)}$.
We can estimate $\Rtilde_{\rm soft}^{(3D)}$ by imposing that the material within
the near wake region contributes more than $80\%$ of the total drag. 
As curvature terms are a pure 2D effect,
the near wake region is the same as in the 2D case. Therefore, $\Rtilde_{\rm soft}^{(3D)}$
satisfies
\begin{equation}
F_{\rm df}^{(3D)}\bigg|_{\Rtilde_{\rm soft}^{(3D)}} = 
5 f_{\rm df}^{(3D)}  \bigg|_{5\Rtilde_{\rm soft}^{(2D)}},
\label{eq:2Dto3D_condition}
\end{equation}
where $f_{\rm df}^{(3D)} (r)$ is the drag force arising from material beyond a distance $r$ 
from the perturber. Following \citet{can13}, we have computed $f_{\rm df}^{(3D)}(r)$
(see top panel of Figure \ref{fig:2Dto3D}), and then obtained ${\mathcal{E}}_{\rm max}^{(3D)}$ 
as a function of  ${\mathcal{E}}_{\rm max}^{(2D)}$ (bottom panel in Figure \ref{fig:2Dto3D}). 
We see that ${\mathcal{E}}_{\rm max}^{(3D)} < {\mathcal{E}}_{\rm max}^{(2D)}$, i. e. the
local approximation in a 3D disk requires even smaller perturbers than in a 2D disk.

For a point-like perfect accretor such as a black hole, the drag force including the 
aerodynamical drag due to accretion is, in the local approximation, 
\begin{equation}
F_{\rm df}^{(3D)}= \frac{\sqrt{8\pi}\Sigma (GM_{p})^{2}}{V^{2}_{\rm rel} H} \ln\left(\frac{7.15H}{R_{\rm acc}}\right)
\end{equation}
\citep{can13}. This formula is very similar to Equation (\ref{eq:canto_softened}) except
the numerical value of the factor in the logarithm, which is larger in the case of a perfect 
accretor, reflecting the fact that accretion contributes to the drag force. 
Therefore, we are certain that the local approximation will be satisfactory
if $R_{\rm acc} < \Rtilde_{\rm soft}^{(3D)}$.

As the accretion radius is given by $R_{\rm acc}\equiv 2GM_{p}/V_{\rm rel}^{2}$, 
its maximum occurs at apocenter, when $V_{\rm rel}$ reaches its minumum value. At apocenter,
$V_{\rm rel}\simeq e\omega a/[2(1+e)^{1/2}]$ and, thus, 
$R_{\rm acc}\simeq 8(1+e) qa/e^{2}$, where we
have used that $GM_{p}=q \omega^{2}a^{3}$.
Hence, the condition $R_{\rm acc} < \Rtilde_{\rm soft}^{(3D)}$ can be cast in terms of $q$ as 
\begin{equation}
q\lesssim \frac{e^{2}{\mathcal{E}}_{\rm max}^{(3D)} h}{8(1+e)}.
\label{eq:upperq}
\end{equation}

For illustration, in the following we discuss some relevant cases; the values used for 
${\mathcal{E}}_{\rm max}^{(3D)}$ are given in Table \ref{table:parameters_epsilon}.
Consider a point-like perturber with $e=0.3$ embedded in a disk with
$\alpha=0$ and $h=0.04$. The local approximation will have an accuracy better
than $20\%$ if $q\lesssim 1.5\times 10^{-4}$,
where we have used ${\mathcal{E}}_{\rm max}^{(3D)}=0.46$ in this case.
It is interesting to compare this upper value
with $q_{\rm crit}$ defined in Section \ref{sec:model_description}. The simulations of \citet{hos07}
indicate that $q_{\rm crit}\gg 10^{-4}$ for $e>0.2$. Therefore,
for $e=0.3$, an accretor with a mass in the range  $1.5\times 10^{-4} < q < q_{\rm crit}$ 
satisfies the type I condition but the local approximation might not be accurate.

Analogously, for $\alpha=0$, $h=0.04$ and $e=0.6$, we have ${\mathcal{E}}_{\rm max}^{(3D)}=0.08$,
and Equation (\ref{eq:upperq}) implies
$q\lesssim 0.8\times 10^{-4}$. For a thicker disk with $h=0.1$, we obtain
$q\lesssim 1.5 \times 10^{-3}$ (again for $\alpha=0$ and $e=0.6$).

\section{Conclusions}
\label{sec:conclusions}

In this paper, we have investigated the quality of the local approximation to estimating
the tidal force acting on low-mass perturbers on eccentric orbits embedded in gaseous
disks. To this aim, we have carried out 2D simulations of perturbers on fixed eccentric orbits
with eccentricities between $0.1$ and $0.6$ in disks with constant aspect ratios ranging
between $0.04$ and $0.1$. In all our simulations, the smoothing length of the perturber is 
larger than the accretion radius.

We find that the local approximation is good if (1) the parameter $X=e/h$ is larger than
$2.5$ so that the perturber moves supersonically relative to the gas and 
(2) the softening radius 
is smaller than a certain threshold value $\Rtilde_{\rm soft}$ so that the force contribution 
of the far-field, which is not well captured in the local approximation, is small. 
Since we are in the regime $R_{\rm acc}<R_{\rm soft}$, an upper value on $R_{\rm soft}$
implies an upper value on $q$.

We have first studied the short-term evolution, that is, when the companion has completed around
$12$ orbits. At those times and 
for an aspect ratio typical in a protoplanetary disk $h=0.04$, the local
approximation can reproduce pretty well both the power and the torque as a function of the orbital phase
for a value of ${\mathcal{E}} \leq 0.15$. The mean values of the power and the torque over one orbit
are well predicted in the local approximation.

On a longer timescale, some models exhibit temporal variations in the torque because
the disk does not reach a periodic configuration in the runtime of our simulations ($\sim 
400$ orbits). These variations occur on a characteristic timescale $\tau_{\rm var}$ of 
$\sim 150$ orbits. In some models, the amplitude of these variations is remarkable.
For instance, for $\alpha=0$ and $R_{\rm soft}=0.024R$, the amplitude of the oscillations is 
comparable to the mean value if $X>12$. In those models that display such large amplitudes, the local approximation still predicts the force during the first stage of the run, at $t \lesssim 15$ orbits, 
but it obviously fails to account for the subsequent changes in the force.
Hence, for those models, the local approximation can
be applied if $q$ is large enough that $\tau_{e}\ll \tau_{\rm var}$.

The amplitude of these variations increases with $\alpha$, $X$ and $R_{\rm soft}$.
Given $\alpha$ and $X$,
the amplitude of the changes in the torque can be reduced by decreasing $R_{\rm soft}$.
By imposing an upper limit on the amplitude of these
torque fluctuations, we have established the threshold value $\Rtilde_{\rm soft}^{(2D)}$
for the local approximation to be faithful.

An extension of the formula for the drag force in the local approximation that incorporates the 
vertical structure of the disk was proposed by \citet{can13}. We have been able to 
determine the validity domain of the 3D local approximation.  We have found the corresponding 
threshold softening radius $\Rtilde_{\rm soft}^{(3D)}$ in the 3D case.

In the case of point-like perturbers, the relevant length is the accretion radius. 
In order for the 3D local approximation to be valid in this case, the accretion radius is required to 
satisfy $R_{\rm acc}\leq \Rtilde_{\rm soft}^{(3D)}$. This condition imposes an upper limit to the value
of $q$. In the case of thin disks ($h\simeq 0.04$) with $0\leq \alpha\leq 1/2$, we 
have found that, for objects with $q\lesssim 10^{-5}$, the 3D local
approximation can be used to determine the orbital evolution in the entire range of orbital
eccentricities considered (i.e. $e \in [0.1, 0.6]$).
This mass range includes the extreme mass-ratios inspirals of BHs in active galactic nuclei 
\citep[e.g.][]{koc11}.
It also includes planetary cores and embryos up to $3$ Earth masses in their 
natal protoplanetary disks.
For thicker disks, the eccentricity range of validity is shifted towards larger values.

\acknowledgments
The author thanks the referee for a thoughtful and constructive report.
The simulations were performed using the computer Tycho (Posgrado de Astrof\'{\i}sica-UNAM,
Instituto de Astronom\'{\i}a-UNAM and PNPC-CONACyT). 
Financial support from PAPIIT project IN111118 is gratefully acknowledged.

\appendix
\section{A. Contribution of the background disk to the power: Mestel disk}
\label{app:mestel}
The unperturbed disk is assumed to be axisymmetric. As a result, the unperturbed disk cannot
create a torque on the orbiting body. However, it can induce a power for bodies with orbital
eccentricity. The power arising from the unperturbed disk is denoted by ${\mathcal{P}}_{d,0}$ and it is given by
\begin{equation}
{\mathcal{P}}_{d,0}=\vecv_{p}\cdot \vecF_{d,0},
\end{equation}
where $\vecv_{p}$ is the velocity of the body and $\vecF_{d,0}$ is the forced exerted on
the body by the unperturbed disk. For concreteness, we focus on a Mestel disk, whose density
decays with $R$ as $\Sigma (R)=\Sigma_{a}a/R$, where $a$ is the semimajor axis and
$\Sigma_{a}$ is the surface density at $R=a$. The gravitational attraction between the unperturbed disk and
a body located at a radius $r_{p}$ is 
\begin{equation}
\vecF_{d,0}= -\frac{2\pi G \Sigma_{0} R_{0} M_{p} }{r_{p}} \hat{\vece}_{r},
\end{equation}
where we have assumed $r_{p}\gg R_{\rm soft}$.
For a body in quasi-Keplerian orbit with semimajor axis $a$ and eccentricity $e$, the power
${\mathcal{P}}_{d,0}$ as a function of $f$ can be written as
\begin{equation}
{\mathcal{P}}_{d,0} = -\frac{2\pi p e q \omega^{3} a^{4} \Sigma_{a}}{\eta^{3}} \sin f \,(1+e \cos f).
\end{equation}
Here we have used that $GM_{p} \simeq q \omega^{2} a^{3}$.
As expected, for a circular orbit ${\mathcal{P}}_{d,0} = 0$ and thus ${\mathcal{P}}_{\rm tot}={\mathcal{P}}_{1}$.
In the following we will focus on eccentric orbits.

\begin{figure*}
\hskip 4.7cm
\includegraphics[width=99mm,height=80mm]{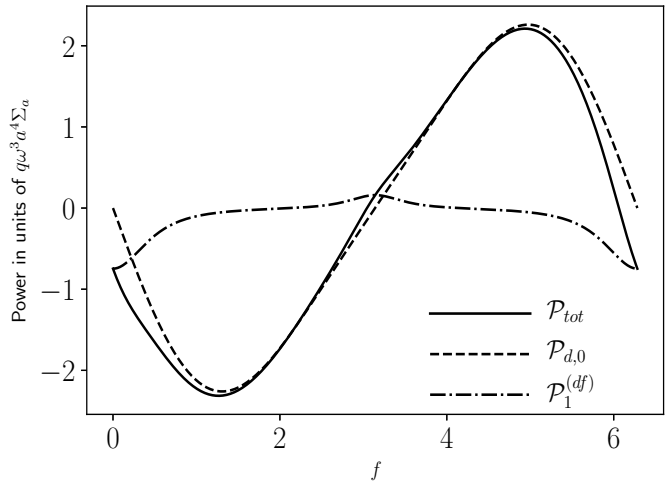}
  \caption{Energy change per unit of time (power) as a function of the true anomaly for an
orbit with $e=0.3$ in a Mestel disk with $h=0.05$. The pertuber has $q=6\times 10^{-5}$ and
$R_{\rm soft}=0.6H$. The dashed line represents the contribution
${\mathcal{P}}_{d,0}$, the dot-dashed line indicates ${\mathcal{P}}^{\rm (df)}_{1}$ and the solid line is ${\mathcal{P}}_{\rm tot}$, i.e. the sum of both contributions.
 }
\vskip 0.75cm
\label{fig:power_background}
\end{figure*}

Figure \ref{fig:power_background} shows ${\mathcal{P}}_{d,0}$, ${\mathcal{P}}^{\rm (df)}_{1}$ 
(from Equation \ref{eq:power_DF})
and ${\mathcal{P}}_{\rm tot}={\mathcal{P}}_{d,0}+{\mathcal{P}}^{\rm (df)}_{1}$ for a Mestel disk with $h=0.05$ and  
a perturber with $q=6\times 10^{-5}$, $e=0.3$ and $R_{\rm soft}=0.6H$. These values are the 
same as in \citep{cre07} to facilitate comparison. It is apparent that the contribution
of ${\mathcal{P}}^{\rm (df)}_{1}$ is disguised if we look at ${\mathcal{P}}_{\rm tot}$.
Although ${\mathcal{P}}^{\rm (df)}_{1}$ appears to be much smaller in amplitude than 
${\mathcal{P}}_{d,0}$, it is the relevant part of the power that
determines the orbital evolution of the particle. In fact, ${\mathcal{P}}_{d,0}$ does not contribute
to the change of the orbital elements because $\mathcal{P}_{d,0}=0$.
It is also important to notice that ${\mathcal{P}}_{d,0}$ depends linearly on $q$, while 
${\mathcal{P}}^{\rm (df)}_{1}$ depends
quadratically. Therefore, the contribution of ${\mathcal{P}}_{d,0}$ relative to ${\mathcal{P}}^{\rm (df)}_{1}$ 
decreases as $q$ increases.

\section{B. The Mach $1$ distance}
\label{app:Mach1}
At apocenter, the local gas rotates at a velocity larger than the perturber. Since the circular
velocity of the gas declines with $R$, there exists a distance in the outer disk (i.e. at $R>[1+e]a$)
beyond which the relative velocity between a patch of gas and the perturber comes subsonic. 
In a disk with constant $h$, the Mach 1 distance $\Delta_{1}$ at apocenter is 
\begin{equation}
\Delta_{1}^{(\rm apo)} = (1+e)a \left[ \frac{1}{(\sqrt{1-e}+h)^{2}} -1\right].
\end{equation} 
At pericenter, the orbiter rotates supersonically with respect to the local gas. However, 
at a distance
\begin{equation}
\Delta_{1}^{(\rm peri)}= (1-e)a \left[1- \frac{1}{(\sqrt{1+e}-h)^{2}} \right]
\end{equation}
interior to the perturber orbit, the relative velocity is $\simeq c_{s}$.
For $e=0.3$ and $h=0.04$, we have that $\Delta_{1}^{(\rm apo)}=0.39a=7.5H_{\rm apo}$ and 
$\Delta_{1}^{(\rm peri)}=0.12a=4.3H_{\rm peri}$, where $H_{\rm apo}$ and $H_{\rm peri}$ are the
scaleheight of the disk at apocenter and pericenter, respectively. This implies that 
the typical scale where the relative motion is supersonic is $\gtrsim 4H$, as long as $e\geq 0.3$.
For $e=0.1$, we find that $\Delta_{1}^{(\rm apo)}=0.43 H_{\rm apo}$ and 
$\Delta_{1}^{(\rm peri)}=0.58 H_{\rm peri}$.


\begin{thebibliography}{}

\bibitem[Artymowicz(1994)]{art94}
Artymowicz, P. 1994, \apj, 423, 581
\bibitem[Baruteau et al.(2014)]{bar14}
Baruteau, C., Crida, A., Paardekooper, S.-J., Masset, F., Guilet, J., Bitsch, B., Nelson, R., Kley, W.,
\& Papaloizou, J. 2014, Protostars and Planets VI, Henrik Beuther, Ralf S. Klessen, Cornelis P. Dullemond, 
and Thomas Henning (eds.), University of Arizona Press, Tucson, 914 pp., p.667-689
\bibitem[Ben\'{\i}tez-Llambay \& Masset(2016)]{ben16}
Ben\'{\i}tez-Llambay, P., \& Masset, F. S. 2016, \apjs, 223, 11
\bibitem[Bernal \& S\'anchez-Salcedo(2013)]{ber13}
Bernal, C. G., \& S\'anchez-Salcedo, F. J. 2013, ApJ, 775, 72
\bibitem[Bitsch \& Kley(2010)]{bit10}
Bitsch, B., \& Kley, W. 2010, \aap, 523, 30
\bibitem[Bitsch \& Kley(2011)]{bit11}
Bitsch, B., \& Kley, W. 2011, \aap, 530, 41
\bibitem[Bitsch et al.(2013)]{bit13}
Bitsch, B., Crida, A., Libert, A.-S., \& Lega, E. 2013, \aap, 555, 124
\bibitem[Cant\'o et al.(2013)]{can13}
Cant\'o, J., Esquivel, A., S\'anchez-Salcedo, F. J., \& Raga, A. C. 2013, \apj, 762, 21
\bibitem[Cresswell \& Nelson(2006)]{cre06}
Cresswell, P., \& Nelson, R. P. 2006, \aap, 450, 833
\bibitem[Cresswell et al.(2007)]{cre07}
Cresswell, P., Dirksen, G., Kley, W., \& Nelson, R. P. 2007, \aap, 473, 329
\bibitem[de Val-Borro et al.(2006)]{dev06}
de Val-Borro, M., Edgar, R. G., Artymowicz, P., et al. 2006, \mnras, 370, 529
\bibitem[Duffell \& Chiang(2015)]{duf15}
Duffell, P. C., \& Chiang, E. 2015, \apj, 812, 94
\bibitem[Fendyke \& Nelson(2014)]{fen14}
Fendyke, S. M., \& Nelson, R. P. 2014, \mnras, 437, 96
\bibitem[Goldreich \& Tremaine(1980)]{gol80}
Goldreich, P., \& Tremaine, S. 1980, \apj, 241, 425
\bibitem[Goldreich \& Sari(2003)]{gol03}
Goldreich, P., \& Sari, R. 2003, \apj, 585, 1024
\bibitem[Grishin \& Perets(2015)]{gri15}
Grishin, E., \& Perets, H. B. 2015, \apj, 811, 54
\bibitem[Hosseinbor et al.(2007)]{hos07}
Hosseinbor, A. P., Edgar, R. G., Quillen, A. C., \& LaPage, A. 2007, \mnras, 378, 966
\bibitem[Just \& Pe\~narrubia(2005)]{jus05}
Just, A., \& Pe\~narrubia, J. 2005, \aap, 431, 861
\bibitem[Kocsis et al.(2011)]{koc11}
Kocsis, B., Yunes, N., \& Loeb, A. 2011, PRD, 84, 024032
\bibitem[Kim \& Kim(2007)]{kim07}
Kim, H., \& Kim, W.-T. 2007, ApJ, 665, 432
\bibitem[Marcy et al.(2005)]{mar05}
Marcy, G., Butler, R. P., Fischer, D., Vogt, S., Wright, J. T., Tinney, C. G., \& Jones, H. R. A. 2005,
Prog. Theo. Physics Supp., 158, 24
\bibitem[Marzari \& Nelson(2009)]{mar09}
Marzari, F., \& Nelson, A. F. 2009, \apj, 705, 1575
\bibitem[Masset(2002)]{mas02}
Masset, F. 2002, \aap, 387, 605
\bibitem[Mills et al.(2019)]{mil19}
Mills, S. M., Horward, A. W., Petigura, E. A., Fulton, B. J., Isaacson, H., \& Weiss, L. M. 2019, \aj, 157, 198
\bibitem[M\"uller et al.(2012)]{mul12}
M\"uller, T. W. A., Kley, W., \& Meru, F. 2012, \aap, 541, 123
\bibitem[Muto et al.(2011)]{mut11}
Muto, T., Takeuchi, T., \& Ida, S. 2011, \apj, 737, 37
\bibitem[Papaloizou(2002)]{pap02}
Papaloizou, J. C. B. 2002, \aap, 388, 615
\bibitem[Papaloizou \& Larwood(2000)]{pap00}
Papaloizou, J. C. B., \& Larwood, J. D. 2000, \mnras, 315, 823
\bibitem[Ragusa et al.(2018)]{rag18}
Ragusa, E., Rosotti, G., Teyssandier, J., Booth, R., Clarke, C. J., \& Lodato, G. 2018, \mnras, 474, 4460
\bibitem[Rein(2012)]{rei12}
Rein, H. 2012, \mnras, 422, 3611
\bibitem[S\'anchez-Salcedo \& Brandenburg(2001)]{san01}
S\'anchez-Salcedo, F. J., \& Brandenburg, A. 2001, MNRAS, 322, 67
\bibitem[S\'anchez-Salcedo et al.(2018)]{san18}
S\'anchez-Salcedo, F. J., Chametla, R. O., \& Santill\'an, A. 2018, \apj, 860, 129
\bibitem[Tamuz et al.(2008)]{tam08}
Tamuz, O., S\'{e}gransan, D., Udry, S., et al. 2008, \aap, 480, L33
\bibitem[Tanaka \& Ward(2004)]{tan04}
Tanaka, H., \& Ward, W. R. 2004, \apj, 602, 388
\bibitem[Udry \& Santos(2007)]{udr07}
Udry, S., \& Santos, N. C. 2007, \araa, 45, 397
\bibitem[Wittenmyer et al.(2007)]{wit07}
Wittenmyer, R. A., Endl, M., Cochran, W. D., \& Levison, H. F. 2007, \aj, 134, 1276
\bibitem[Xie et al.(2016)]{xie16}
Xie, J.-W., Dong, S., Zhu, Z., et al. 2016, Proceedings of the National Academy of Sciences
of the United States of America, 113, 11431

\end{thebibliography}
\end{document}